\documentclass[sigconf,screen]{acmart}

\usepackage[english]{babel}

\usepackage{booktabs}
\usepackage{colortbl}
\usepackage{enumitem}
\usepackage{float}
\usepackage{graphicx}
\usepackage[utf8]{inputenc}
\usepackage{multirow}
\usepackage[group-separator={,},per-mode=symbol]{siunitx}
\usepackage{subcaption}
\usepackage{textcomp}
\usepackage{xcolor}
\usepackage{xspace}
\usepackage[normalem]{ulem}

\usepackage{hyperref}

\usepackage[super]{nth}

\newif\ifEditMode

\EditModefalse

\AtBeginDocument{%
  \providecommand\BibTeX{{%
    \normalfont B\kern-0.5em{\scshape i\kern-0.25em b}\kern-0.8em\TeX}}}

\hypersetup{
  bookmarksnumbered=true,
  bookmarksopen=true,
  citecolor=blue,
  linkcolor= blue,
  urlcolor=blue}

\newcommand{\term}[1]{\emph{#1}}
\newcommand{\newterm}[1]{\emph{#1}}
\newcommand{\stress}[1]{\textit{#1}}

\definecolor{maroon}{rgb}{0.5, 0.0, 0.0}
\newcommand{\attention}[1]{\textcolor{maroon}{\textbf{#1}}}%

\newcommand{\tsup}[1]{\ensuremath{^{\textrm{#1}}}}

\newcommand{\parai}[1]{\vspace{0.03in}\noindent{\textit{#1}}\quad}
\newcommand{\paraib}[1]{\vspace{0.03in}\noindent{\textit{\textbf{#1}}}\quad}
\newcommand{\point}{{\noindent}~$\blacktriangleright$\,}
\newcommand{\thead}[1]{\textbf{\textit{#1}}}
\newcommand{\figcap}[1]{\caption{\textit{#1}}}
\newcommand{\sfigcap}[1]{\caption{\textit{#1}}}
\newcommand{\tabcap}[1]{\vspace*{5pt}\caption{\textit{#1}}}

\newcommand{\usd}[1]{\SI{#1}{\second}}

\newcommand{\uMB}[1]{\SI{#1}{MB}}

\newcommand{\uBTC}[1]{\SI{#1}{BTC}}
\newcommand{\uTxFee}[1]{\SI{#1}{BTC{\per}KB}}
\newcommand{\pow}[0]{proof-of-work\xspace}
\newcommand{\dsa}[0]{$\mathcal{A}$\xspace}
\newcommand{\dsb}[0]{$\mathcal{B}$\xspace}
\newcommand{\dsc}[0]{$\mathcal{C}$\xspace}

\newcommand{\gbt}{\texttt{GetBlockTemplate}\xspace}
\newcommand{\GBT}{GBT\xspace}
\newcommand{\mpool}{Mempool\xspace}

\newcommand{\feeunit}{BTC/KB}
\newcommand{\DefTxFee}[0]{\SI{1}{sat{\per}byte}}

\newlength{\onecolgrid}
\newlength{\twocolgrid}
\newlength{\threecolgrid}
\newlength{\fourcolgrid}
\ifEditMode
\newcommand{\editlabel}[1]{\raisebox{.35ex}{\tiny \scshape[#1]}}

\newcommand{\REVISIT}[1]{%
  \textcolor[HTML]{984ea3}{\editlabel{Revisit} #1}}\xspace{}

\newcommand{\PENDING}[1]{%
  \textcolor[HTML]{e67e22}{$\cdots$ \editlabel{Pending} \textit{#1} $\cdots$}\xspace{}}

\newcommand{\TODO}[1]{\textcolor[HTML]{e41a1c}{{(#1)}}}

\newcommand{\FIXME}[2]{%
  \textcolor[HTML]{c0392b}{\editlabel{FIXME(#1)} \textbf{#2}}}
\newcommand{\MISSING}[1]{\textcolor[HTML]{de2d26}{\textbf{#1}}}

\newcommand{\REMOVE}[1]{%
  {\small \textcolor[HTML]{a65628}{#1\\--- Consider deleting.}}}

\newcommand{\krishna}[1]{\textcolor{blue}{#1}}
\newcommand{\patrick}[1]{\textcolor{brown}{#1}}

\else
\makeatletter

\newcommand{\REVISIT}[1]{#1}

\newcommand{\PENDING}[1]{\@bsphack\@esphack}
\newcommand{\TODO}[1]{\@bsphack\@esphack}
\newcommand{\FIXME}[2]{\@bsphack\@esphack}
\newcommand{\MISSING}[1]{\@bsphack\@esphack}
\newcommand{\REMOVE}[1]{\@bsphack\@esphack}

\newcommand{\krishna}[1]{\@bsphack\@esphack}
\newcommand{\patrick}[1]{\@bsphack\@esphack}

\makeatother
\fi

\acmYear{2021}\copyrightyear{2021}
\setcopyright{rightsretained}
\acmConference[IMC '21]{ACM Internet Measurement Conference}{November 2--4, 2021}{Virtual Event, USA}
\acmBooktitle{ACM Internet Measurement Conference (IMC '21), November 2--4, 2021, Virtual Event, USA}
\acmPrice{}
\acmDOI{10.1145/3487552.3487823}
\acmISBN{978-1-4503-9129-0/21/11}



\begin{document}
\title{Selfish \& Opaque Transaction Ordering in the Bitcoin Blockchain: The Case for Chain Neutrality}

\author[J. Messias]{Johnnatan Messias}
\email{johnme@mpi-sws.org}
\affiliation{%
  \institution{MPI-SWS}
  \country{Germany}
}

\author[M. Alzayat]{Mohamed Alzayat}
\email{alzayat@mpi-sws.org}
\affiliation{%
  \institution{MPI-SWS}
  \country{Germany}
}

\author[B. Chandrasekaran]{Balakrishnan Chandrasekaran}
\email{b.chandrasekaran@vu.nl}
\affiliation{%
  \institution{Vrije Universiteit Amsterdam}
  \country{Netherlands}
}
\author[K. P. Gummadi]{Krishna P. Gummadi}
\email{gummadi@mpi-sws.org}
\affiliation{%
  \institution{MPI-SWS}
  \country{Germany}
}

\author[P. Loiseau]{Patrick Loiseau}
\email{patrick.loiseau@inria.fr}
\affiliation{%
  \institution{Univ. Grenoble Alpes, Inria, CNRS, Grenoble INP, LIG}
  \country{France}
}

\author[A. Mislove]{Alan Mislove}
\email{amislove@ccs.neu.edu}
\affiliation{%
  \institution{Northeastern University}
  \country{USA}
}

\begin{abstract}
    Most public blockchain protocols, including the popular Bitcoin and Ethereum blockchains, do not formally specify the order in which miners should select transactions from the pool of pending (or uncommitted) transactions for inclusion in the blockchain.
Over the years, informal conventions or ``norms'' for transaction ordering have, however, emerged via the use of shared software by miners, e.g., the \texttt{GetBlockTemplate (GBT)} mining protocol in Bitcoin Core.
Today, a widely held view is that Bitcoin miners prioritize transactions based on their offered ``transaction fee-per-byte.''
Bitcoin users are, consequently, encouraged to increase the fees to accelerate the commitment of their transactions, particularly during periods of congestion.
In this paper, we audit the Bitcoin blockchain and present statistically significant evidence of mining pools deviating from the norms to accelerate the commitment of transactions for which they have (i) a selfish or vested interest, or (ii) received dark-fee payments via opaque (non-public) side-channels. 
As blockchains are increasingly being used as a record-keeping substrate for a variety of decentralized (financial technology) systems, our findings call for an urgent discussion on defining neutrality norms that miners must adhere to when ordering transactions in the chains.
Finally, we make our data sets and scripts publicly available.


\end{abstract}

\maketitle

\section{Introduction} \label{sec:introduction}

%

At its core, a blockchain is an 
append-only 
list of cryptographically linked records of transactions called ``blocks.''   
In 
public blockchains such as Bitcoin~\cite{Nakamoto-WhitePaper2008} and Ethereum~\cite{Wood@Ethereum}, any user can broadcast a transaction to be included in the blockchain.
Participants, called miners, include (or confirm) the issued transactions in a new block and extend the blockchain by solving a cryptographic puzzle.  
Many blockchains are maintained in a decentralized manner by a peer-to-peer (P2P) network of nodes that follow a well-defined protocol (i.e., ground rules) for validating new blocks.
For example, the protocol for maintaining the Bitcoin ledger, laid down by Nakamoto in 2008, is based on a proof-of-work (PoW)
scheme~\cite{Nakamoto-WhitePaper2008}. 
Noticeably absent from Bitcoin and other decentralised blockchain protocols is the requirement of any a-priori trust between the users issuing transactions, the miners confirming transactions, and the P2P nodes maintaining the blockchain.

Decentralized blockchains, without any notion of trusted entities, have not only been used to implement cryptocurrencies, but are increasingly being adopted as a substrate for a variety of decentralized financial applications (smart contracts) such as exchanges~\cite{Daian@S&P20,UniswapDEX}, lending~\cite{Qin@FC21,Perez@FC21}, and auctions~\cite{NFTs}. 
Despite their widespread use in ordering critical applications~\cite{Mccorry@FC17,Daian@S&P20,pilkington2016blockchain,kharif2017cryptokitties,UniswapDEX,Perez@FC21}, blockchain protocols formally specify \stress{neither} the manner by which miners should select transactions for inclusion in a new block from the set of all available transactions, \stress{nor} the order in which they should be included in the block.
While informal conventions or norms for prioritizing transactions exist, to our knowledge, no one has systematically verified if these norms are being followed by miners in practice.
In this paper, we present an in-depth analysis of transaction prioritization by Bitcoin miners.

Bitcoin is the largest cryptocurrency in the world, with a market capitalization
of over \$742.6B as of May 2021~\cite{CoinMarketCap-URL2021}.
It has been observed that the increasing volume of Bitcoin transactions issued introduces
\stress{congestion} among transactions for confirmation~\cite{Kuzmanovic-QUEUE2019}:
Due to size limits on Bitcoin blocks, at any time, there may be more transactions than can be immediately committed or confirmed\footnote{We use the terms `confirmation' and `commit' interchangeably to refer to the inclusion of a transaction in a block.} in the next block.
Unconfirmed transactions must, consequently, wait for their ``turn'' to be included in subsequent blocks, thereby introducing \stress{delays}.
So the order in which miners choose transactions for inclusion in a new block crucially determines how long individual transactions (e.g., currency transfers) are delayed.
Worse, some transactions may be \stress{conflicting}, meaning at most one of the transactions can be included in the blockchain; for such transactions, the order in which a miner chooses to include transactions will determine the ultimate state of the system.

%
%

\begin{figure}[tb]
    \centering
    \includegraphics[width={\onecolgrid}]{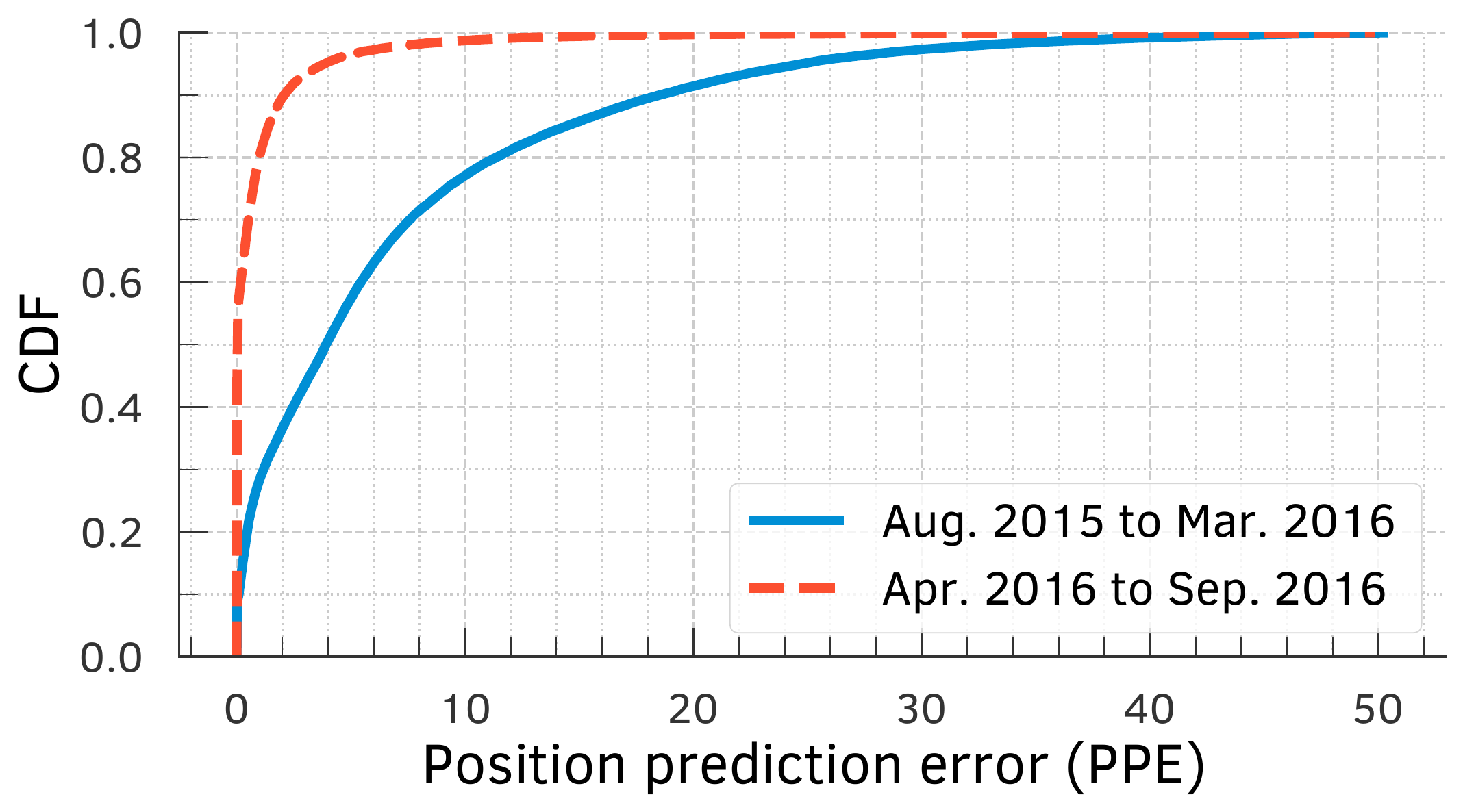}
    \figcap{CDF of the error in predicting where a transaction would be positioned or ordered within a block according to the greedy fee-rate-based norm. Bitcoin Core code shifted completely to the fee-rate-based norm starting April 2016: Transaction ordering in Bitcoin closely tracks the fee-rate-based norm from April 2016, but differs significantly from it prior to April 2016 when a different norm was in place.}
    \label{fig:different-norms-in-btc}
\end{figure}

The conventional wisdom today is that many miners follow the prioritization norms, implicitly, by using widely shared blockchain software like the Bitcoin Core~\cite{BitcoinCore-2021,CoinDance-2021}.
Then, in Bitcoin, the presumed ``norm'' is that miners prioritize a transaction for inclusion based on its offered \stress{fee-rate} or fee-per-byte, which is the transaction's fee divided by the transaction's size in bytes.
We show evidence of this presumed norm in Figure~\ref{fig:different-norms-in-btc}.
The norm is also justified as ``incentive compatible'' because miners wanting to maximize their rewards, i.e., fees collected from all transactions packed into a size-limited block, would be incentivized to include preferentially transactions with higher fee-rates.
Assuming that miners follow this norm, Bitcoin users are issued a crucial recommendation: To accelerate the confirmation of a transaction, particularly during periods of congestion, they should increase the transaction transaction fees.
We show that miners are, however, free to deviate from this norm and such norm violations cause irreparable economic harm to users.

In this paper, we perform an extensive empirical audit of the miners' behavior to check whether they conform to the norms.\footnote{
We use the terms ``miners,'' ``mining pool operators (MPOs),'' and ``mining pools'' interchangeably throughout this paper.}
%
%
At a high-level, we find that transactions are indeed primarily prioritized according to the assumed norms. 
We also, nevertheless, offer evidence of a non-trivial fraction of priority-norm violations amongst confirmed transactions.
An in-depth investigation of these norm violations uncovered many highly troubling mis-behaviors by miners. 
Specifically, we present two key findings.

\indent{}\point{}
Multiple large mining pools tend to {\it selfishly prioritize} transactions in which they have a vested interest; e.g., transactions in which payments are made from or to wallets owned by the mining pool operators. Some even {\it collude} with other large mining pools to prioritize their transactions.

\indent{}\point{}
Many large mining pools accept additional {\it dark (opaque) fees} to accelerate transactions via non-public side-channels (e.g., their websites). Such dark-fee transactions violate an important, but unstated assumption in blockchains that confirmation fees offered by transactions are transparent and equal to all miners.


While some of the above miner misbehaviors have been speculated in prior work~\cite{Kelkar@CRIPTO20,Kursawe@AFT20}, to the best of our knowledge, our work is the first to offer a strong empirical evidence of such miner misbehaviors in practice. 
In the process, we have developed robust tests to detect miner misbehaviors in the Bitcoin blockchain. 
%
We view the design of these tests as an important contribution of independent interest to researchers auditing blockchains.

Our findings have important implications for both Bitcoin users and miners. 
Specifically, when setting fees for their transactions, Bitcoin users (i.e., through their wallet software) assume that the fees offered by all their competing transactions are fully transparent---our findings contradict this assumption.
Similarly, when transactions offer different confirmation fees to different miners, it raises significant unfairness concerns.
Finally, the collusion we uncovered between mining pools exacerbates the growing concerns about the concentration of hash rates amongst a small number of miners~\cite{Gervais@CCS-16,bahack2013theoretical}.
We release the data sets and the scripts used in our analyses to facilitate others to reproduce our results~\cite{Messias-DataSet-Code-2021}.
%

\section{Background}\label{sec:background}

A Bitcoin user or client issues transactions that move currency from one or more
\newterm{wallets} (i.e., addresses) owned by the client to another.
\newterm{Miners}, who are a~subset of these users, validate the transactions and
include them in a \newterm{block}.
A~block is a set of zero\footnote{Miners can mine an ``empty'' block
without including any transaction in it.} or more transactions in addition to
the \newterm{Coinbase} transaction, which moves the rewards to the miner's
wallet.
Until these transactions are included in a block, they remain \newterm{unconfirmed}.
Miners create a block by including such unconfirmed transactions and solving a
cryptographic puzzle that includes, among other things, a hash of the most
recent block mined in the network.
The chain of cryptographic hashes linking each block to an ancestor all the way
to the initial (or \newterm{genesis}) block~\cite{blockchain-2009} constitutes
the blockchain.

Miners are rewarded for their work in two ways.
First, miners reap a block reward upon mining a block.
Second, miners also collect fees, if any, from each transaction; fees are
included by users to incentivize the miners to commit their transaction.
We refer to the software implementation (along with the hardware) used by a miner as a \newterm{node}.
A~node allows a miner to receive broadcasts of transactions and blocks from
their peers, validate the data, and mine a block.
Nodes queue the unconfirmed transactions received via broadcasts in an in-memory
buffer, called the \newterm{\mpool}, from where they are dequeued for inclusion
in a block.
One can also configure the node to skip mining and simply use it as an
observer.

\subsection{Transaction prioritization norms}\label{sec:prelim}

A crucial detail absent in the design of a \pow{} blockchain
per~\cite{Nakamoto-WhitePaper2008} is any notion of a formal specification of
transaction prioritization.
Said differently, Nakamoto's design does not formally specify how miners should
select a set of candidate transactions for confirmation from all available unconfirmed
transactions.
Notwithstanding this shortcoming, ``norms'' have originated from miners' use
of a shared software implementation:
Miners predominantly use the Bitcoin Core~\cite{BitcoinCore-2021} software for
communicating with their peers (e.g., to advertise blocks and learn about new
unconfirmed transactions) and reaching a consensus regarding the chain.
%

Of particular note in the popular Bitcoin core’s implementation is the
\texttt{GetBlockTemplate (GBT)} mining protocol, implemented by the Bitcoin
community around February 2012.\footnote{%
Even within mining pools, the widely used Stratum protocol internally uses the \gbt mechanism~\cite{Stratum-v1-2021}.}
\gbt{} rank orders transactions based on the fee-per-byte (i.e., transaction
fees normalized by the transaction's size) metric~\cite{GBT-Bitcoin-2019}.

The term \stress{size}, here and in the rest of the paper, refers to
\newterm{virtual} size, each unit of which corresponds to four \newterm{weight
units} as defined in the Bitcoin improvement proposal
BIP-141~\cite{Lombrozo-BIP141-2015}.
The predominant use of GBT (through the use of Bitcoin core) by miners coupled
with the fact that GBT is maintained by the Bitcoin community
\stress{implicitly} establishes two norms.
A~third norm stems from a configuration parameter of the Bitcoin core
implementation.
We now elucidate these three norms.

\textbf{I.}
\stress{When mining a new block, miners select transactions for inclusion, from
the \mpool{}, based solely on their fee-rates.}

\textbf{II.}
\stress{When constructing a block, miners order (place) higher fee-rate transactions before lower fee-rate transactions.}

\textbf{III.}
\stress{Transactions with fee-rate below a minimum threshold are ignored and never committed to the blockchain.}

The GBT protocol implementation in Bitcoin core is the source of the first two norms.
GBT's rank ordering determines both which set of transactions are selected for inclusion (from the \mpool{}) and in what order they are placed within a block.
GBT dictates that a transaction with higher fee-per-byte \stress{will} be
selected before all other transactions with a lower fee-per-byte.
It also stipulates that within a block a transaction with the highest fee-per-byte appears first, followed by next highest fee-per-byte, and so on.

The third norm stems from the fee-per-byte threshold configuration parameter.
Bitcoin core, by design, will not accept any transaction with fee-rate below
this threshold, essentially filtering out low-fee-rate transactions from even
being accepted into the \mpool{}.
The default (and recommended) value for this configurable threshold is set to
$\DefTxFee{}$.\footnote{One Bitcoin (BTC) is equal to $10^{8}$ satoshi (sat).}

\subsection{Related Work} \label{sec:related-work}


A few recent papers proposed solutions to enforce that transaction ordering follows a certain norm, mostly based on statistical tests of potential deviations \cite{Orda2019,Asayag18a,lev2020fairledger}. These work were, however, mostly of theoretical nature in that they did not contain empirical evidence of deviation by miners, but rather assumed that miners might deviate. Prior efforts also proposed consensus algorithms to guarantee fair-transaction selection~\cite{baird2016swirlds,Kursawe@AFT20,Kelkar@CRIPTO20}. Kelkar \textit{et al.}~\cite{Kelkar@CRIPTO20} proposed a consensus property called \textit{transaction order-fairness} and a new class of consensus protocols called \textit{Aequitas} to establish fair-transaction ordering in addition to also providing consistency and liveness. A number of prior work focused on enabling miners to select transactions. For instance,  SmartPool~\cite{Luu2017} gave transaction selection back to the miners. Similarly, an improvement of Stratum, a well-used mining protocol, allows miners to select their desired transaction set through negotiation with a mining pool~\cite{Stratum-2021}. All these prior work are, again, mostly of theoretical nature. In contrast, our study provides empirical evidence of deviation from the norm by miners in the current Bitcoin system.


\begin{table*}[tb]

  \begin{center}
   \small
    \tabcap{Bitcoin data sets (\dsa and \dsb) used for testing miners' adherence to transaction-prioritization norms and (\dsc) for investigating the behaviour of mining pool operators}\label{tab:datasets}
    \begin{tabular}{rrrr}
      \toprule
      \thead{Attributes} & \thead{Data set \dsa{}} & \thead{Data set \dsb{}} & \thead{Data set \dsc{}}\\
      \midrule
      \textit{Time span} & Feb. $20\tsup{th}$ -- Mar. $13\tsup{th}$, 2019 & Jun. $1\tsup{st}$ -- $30\tsup{th}$, 2019 & Jan. $1\tsup{st}$ -- Dec. $31\tsup{st}$, 2020\\
      \textit{Block height} & \num{563833} -- \num{566951} & \num{578717} -- \num{583236} & \num{610691} -- \num{663904} \\
    \textit{Number of blocks} & $\num{3119}$ & $\num{4520}$ & $\num{53214}$ \\
      \textit{Count of transactions issued} & $\num{6816375}$ & $\num{10484201}$ & $\num{112489054}$\\
      \textit{Percentage of CPFP-transactions} & $26.45\%$ & $23.17\%$ & $19.11\%$ \\
      \textit{Count of empty-blocks} & \num{38} & \num{18} & \num{240}\\
      \bottomrule
    \end{tabular}
  \end{center}
\end{table*}

\begin{figure*}[tb]
  \centering
  \begin{subfigure}[b]{\threecolgrid}
      \includegraphics[width={\textwidth}]{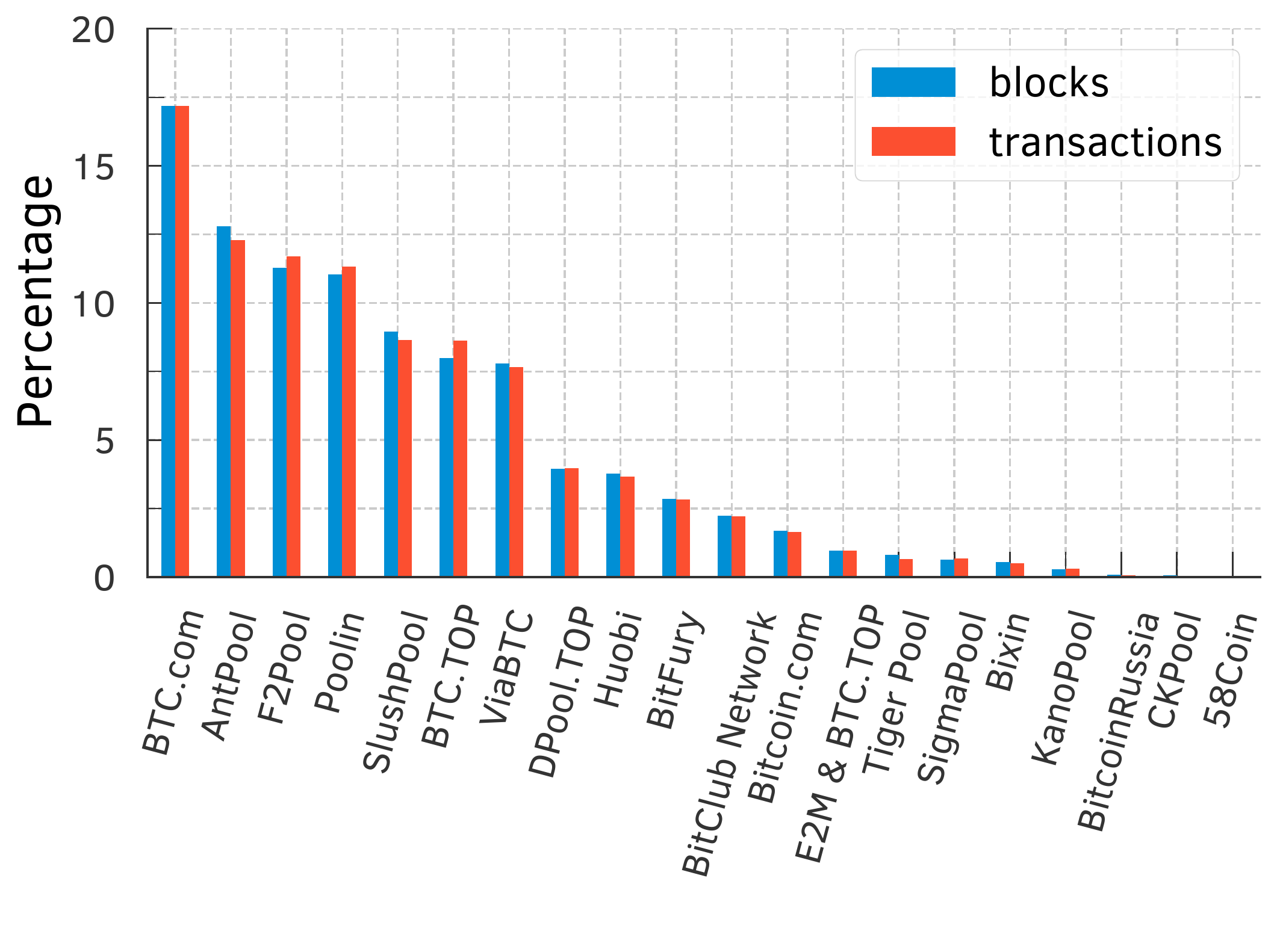}
      \sfigcap{Data set \dsa{}}\label{fig:dist-txs-blks-dataset-a}
    \end{subfigure}
    \begin{subfigure}[b]{\threecolgrid}
      \includegraphics[width={\textwidth}]{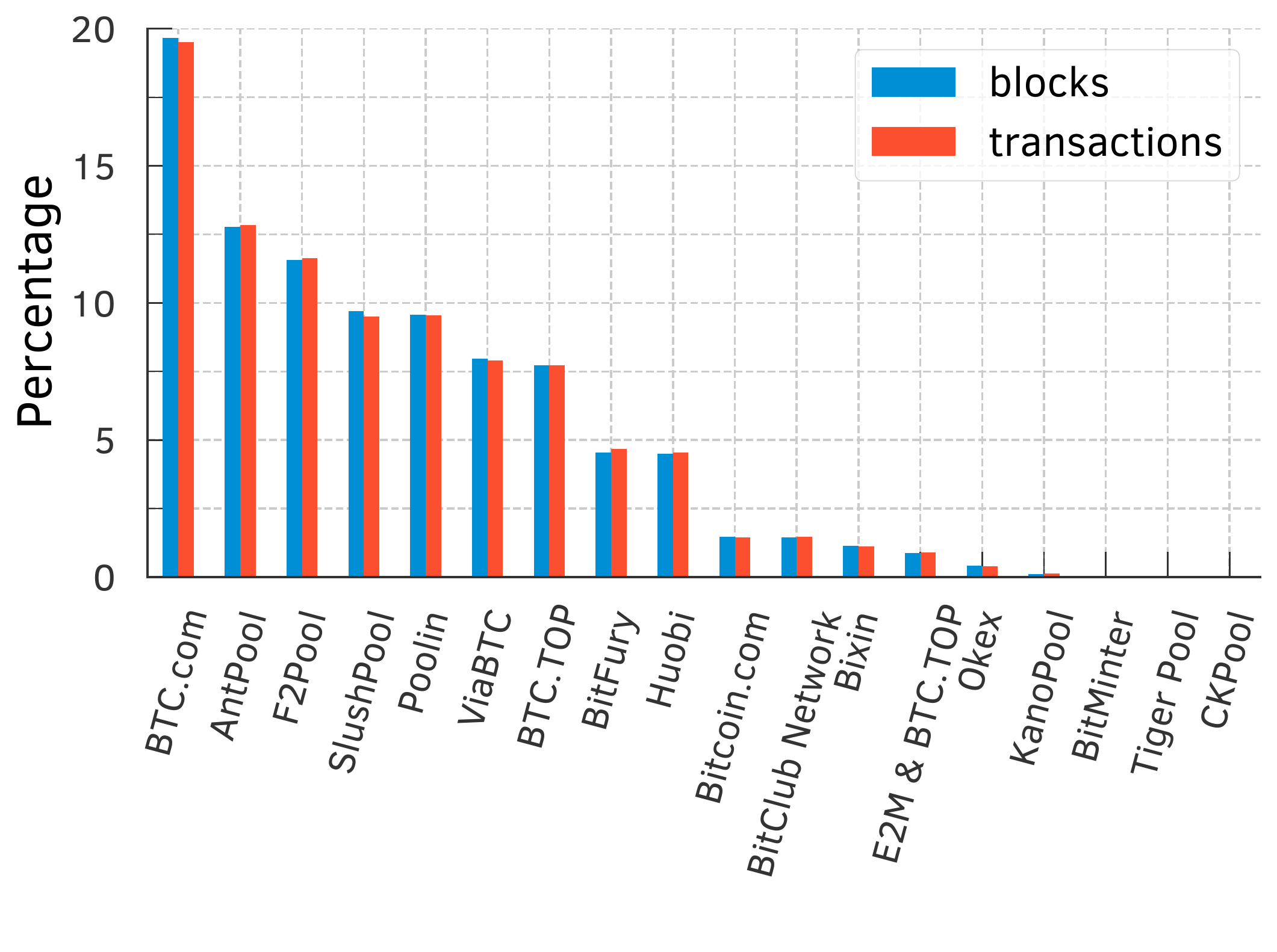}
      \sfigcap{Data set \dsb{}}\label{fig:dist-txs-blks-dataset-b}
    \end{subfigure}
    \begin{subfigure}[b]{\threecolgrid}
      \includegraphics[width={\textwidth}]{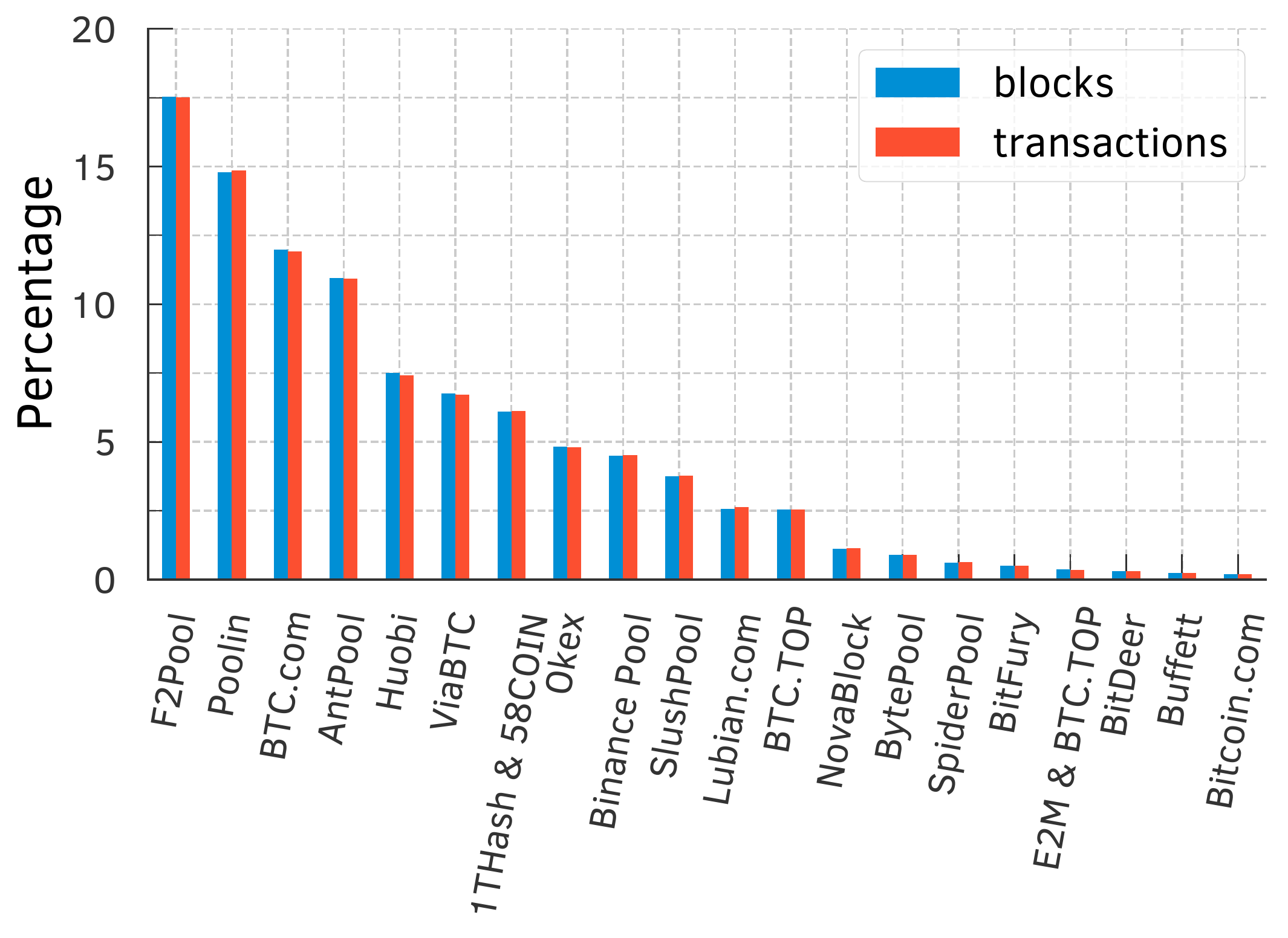}
      \sfigcap{Data set \dsc{}}\label{fig:dist-txs-blks-dataset-c}
    \end{subfigure}
    \figcap{Distribution of blocks mined and transactions confirmed by the top-20 
      MPOs in data sets \dsa{}, \dsb{}, and \dsc{}. Their combined normalized hash-rates account for 94.97\%, 93.52\%, and 98.08\% of all blocks mined in data set \dsa, \dsb, and \dsc, respectively.}\label{fig:dist-tx-blks-dataset-all}
\end{figure*}

Fairness issues have been studied in blockchain from the point of view of miners. Pass \textit{et al.}~\cite{Pass@PODC17} proposed a fair blockchain where transaction fees and block rewards are distributed fairly among miners, decreasing the variance of mining rewards. Other studies focused on the security issues showing that miners should not mine more blocks than their ``fair share''~\cite{Eyal-CACM2018} and that mining rewards payout is centralized in mining pools and therefore unfairly distributed among their miners~\cite{Romiti2019ADD}. Chen \textit{et al.}~\cite{Chen@AFT19} studied the allocation of block rewards on blockchains showing that Bitcoin's allocation rule satisfies some properties. It does not, however, hold when miners are not risk-neutral, which is the case for Bitcoin.
In contrast to these prior work, this paper touches upon fairness issues from the viewpoint of transaction issuers and not miners. 

There is a vast literature on incentives in mining. Most of it, however, considers only block rewards~\cite{Romiti2019ADD,Chen@AFT19,Eyal-CACM2018,Pass_Seeman_Shelat_2017,Zhang_Preneel_2019,sompolinsky2015secure,Kiayias@EC16,Fiat@EC19,Goren@EC19,Noda@EC20}. 
As the block reward halves every four years, some recent work focused on analyzing how the incentives will change when transaction fees dominate the rewards. Carlsen \textit{et al.}~\cite{Carlsten@CCS16} showed that having only transaction fees as incentives will create instability. Tsabary and Eyal~\cite{Tsabary@CCS18} extended this result to more general cases including both block rewards and transaction fees. Easley \textit{et al.}~\cite{Easley19a} proposed a general economic analysis of the system and its welfare with various types of rewards. Those prior work, however, assume that miners follow a certain norm for transaction selection and ordering (mostly the fee-rate norm) and look at miners' incentives in terms of how much compute power to exert and when (or some equivalent metric). 
There are also prior studies on the security issues of having transaction fees as the prime miners' incentive~\cite{Carlsten@CCS16,Li@IV18}; and a vast literature on the security of blockchains more generally (e.g., \cite{Gencer-FC2018,Karame-CCS2016,Vasek-FC2014}). Again, however, these studies focus on miners' incentives to mine and not on transaction ordering; for the latter, they assume that miners follow a norm. These prior studies are, hence, somewhat orthogonal to our work.

Only a few recent work touched upon the issue of how miners select and order transactions, and how this is interlaced with how the fees are set.
Lavi~\textit{et al.}~\cite{Lavi-WWW2019} and Basu~\textit{et al.}~\cite{Basu-CoRR2019} highlighted the inefficiencies in the existing transaction fee-setting mechanisms and proposed alternatives. They showed that miners might not be trustworthy, but without providing empirical evidence. 
Siddiqui \textit{et al.}~\cite{Siddiqui@AAMAS20} showed through simulations that, with transaction fees only as incentives, miners would have to select transactions greedily, increasing the latency for most of the transactions. They proposed an alternative selection mechanism and performed numerical simulations on it.
Our work takes a complementary approach: We analyze empirical evidence of miners deviations from the transaction ordering norm in the current ecosystem. 
We also empirically analyze existing collusion at the level of transaction inclusion.


To the best of our knowledge, our study is the first of its kind---showing empirical evidence of norm violations in Bitcoin---and our results help motivate the theoretical studies mentioned above.


\section{Data Sets} \label{sec:datasets}

To understand the importance of transaction ordering to users and 
to investigate when and how miners violate the
transaction prioritization ``norms,'' we resort to an empirical, data-driven
approach.
Below, we briefly describe three different data sets that we curated from
Bitcoin
and
highlight how we use the data sets in different analyses in the rest of the paper.

\paraib{Data set \dsa{}.}
To check miners' compliance to prioritization norms in Bitcoin, we analyzed all
transactions and blocks issued in Bitcoin over a three-week time frame from
February 20 through March 13, 2019 (see Table~\ref{tab:datasets}).
We obtained the data by running a \term{full} node, a Bitcoin software that
performs nearly all operations of a miner (e.g., receiving broadcasts of
transactions and blocks, validating the data, and re-broadcasting them to peers)
with the exception of mining.
The data set contains a set of periodic \term{snapshots}, recorded once per
$15$ seconds for the entire three-week period, where each snapshot captures the state
of the full node’s \mpool{}.
%
%
We plot the distribution of the count of blocks and transactions mined by the top-20 MPOs for data set \dsa in Figure~\ref{fig:dist-txs-blks-dataset-a}. If we rank the MPOs in data set \dsa by the number of blocks ($B$) mined (or, essentially, the
approximate hashing capacity $h$), the top five MPOs turn out to be BTC.com ($B$: $\num{536}$; $h$: $17.18\%$), AntPool ($B$: $\num{399}$; $h$: $12.79\%$), 
F2Pool ($B$: $\num{352}$; $h$: $11.29\%$), Poolin ($B$: $\num{344}$; $h$: $11.03\%$), and SlushPool ($B$: $\num{279}$; $h$: $8.94\%$).
We use this data for checking whether miners adhere to prioritization norms when selecting transactions for confirmation or inclusion in a block (\S\ref{sec:prioritization-norms}).

\paraib{Data set \dsb{}.}
Differences in configuration of the Bitcoin software may subtly affect the inferences drawn from \dsa{}.
A full node connects to \(8\) peers, for instance, in the default configuration,
and increasing this number may reduce the likelihood of missing a transaction
due to a “slow” peer.
The default configuration also imposes a minimum fee-rate threshold of
$\DefTxFee$ for accepting a transaction.
We instantiated, hence, another full node to expand the scope of our data collection.
We configured this second node, for instance, to connect to as many as \(125\) peers.
We also removed the fee-rate threshold to accept even zero-fee transactions.
\dsb{} contains \mpool{} snapshots of this full node, also recorded once per \usd{15},
for the entire month of June 2019 (refer Table~\ref{tab:datasets}). We notice that $99.7\%$ of the transactions received by our \mpool{} were included by miners.
Figure~\ref{fig:dist-txs-blks-dataset-b} shows the distribution of the count of blocks and transactions mined by the top-20 MPOs for data set \dsb. The top five MPOs are BTC.com ($B$: $\num{889}$; $h$: $19.67\%$), AntPool ($B$: $\num{577}$; $h$: $12.77\%$), 
F2Pool ($B$: $\num{523}$; $h$: $11.57\%$), SlushPool ($B$: $\num{438}$; $h$: $9.69\%$), and Poolin ($B$: $\num{433}$; $h$: $9.58\%$).
As in the case of \dsa, we use this data set in \S\ref{sec:prioritization-norms}.
%



\begin{figure*}[tbh]
    \centering
    \begin{subfigure}[b]{\threecolgrid}
        \includegraphics[width={\textwidth}]{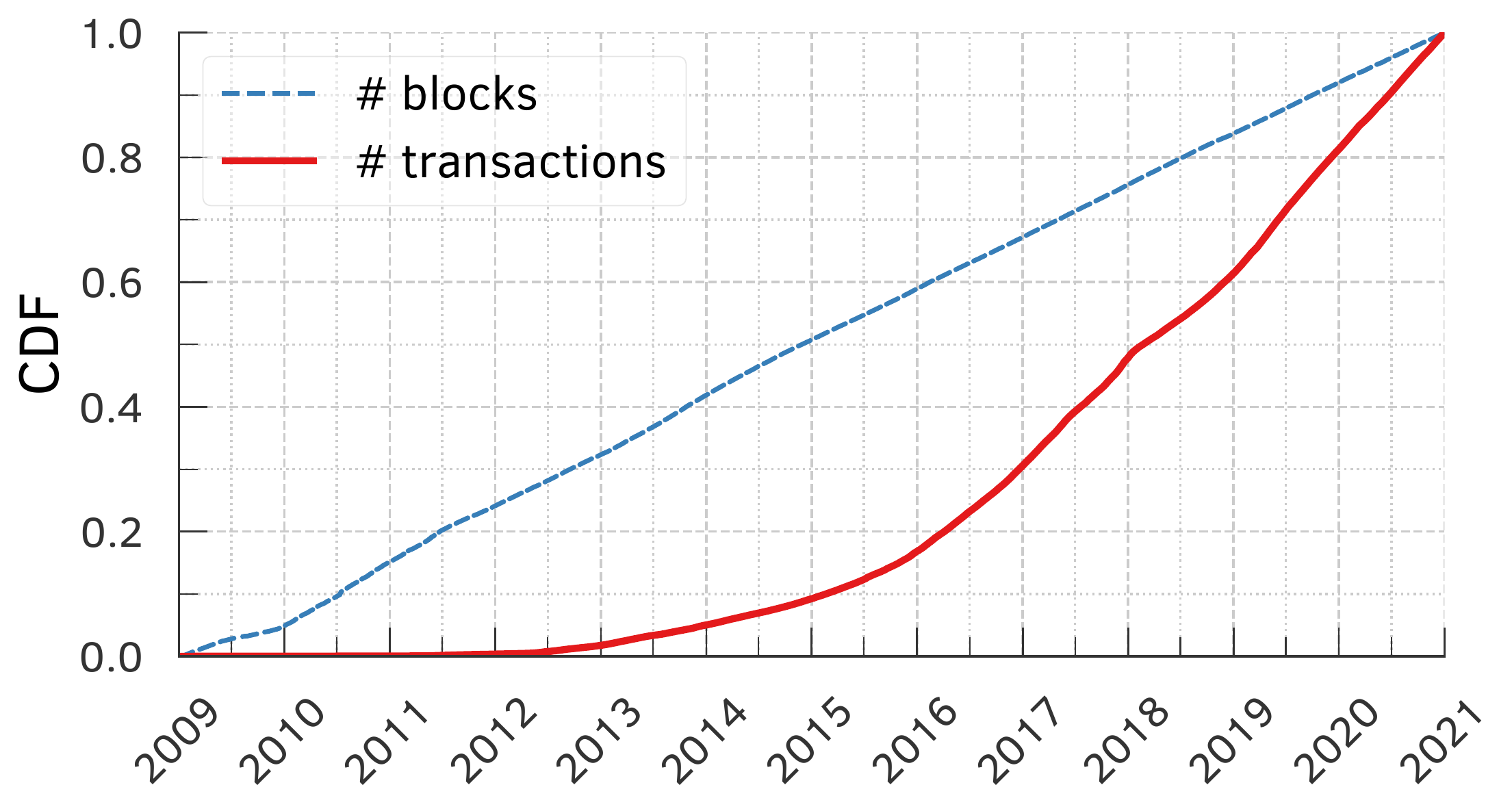}
        \sfigcap{}\label{fig:cdf-tx-blks-btc}
    \end{subfigure}
    \begin{subfigure}[b]{\threecolgrid}
        \includegraphics[width={\textwidth}]{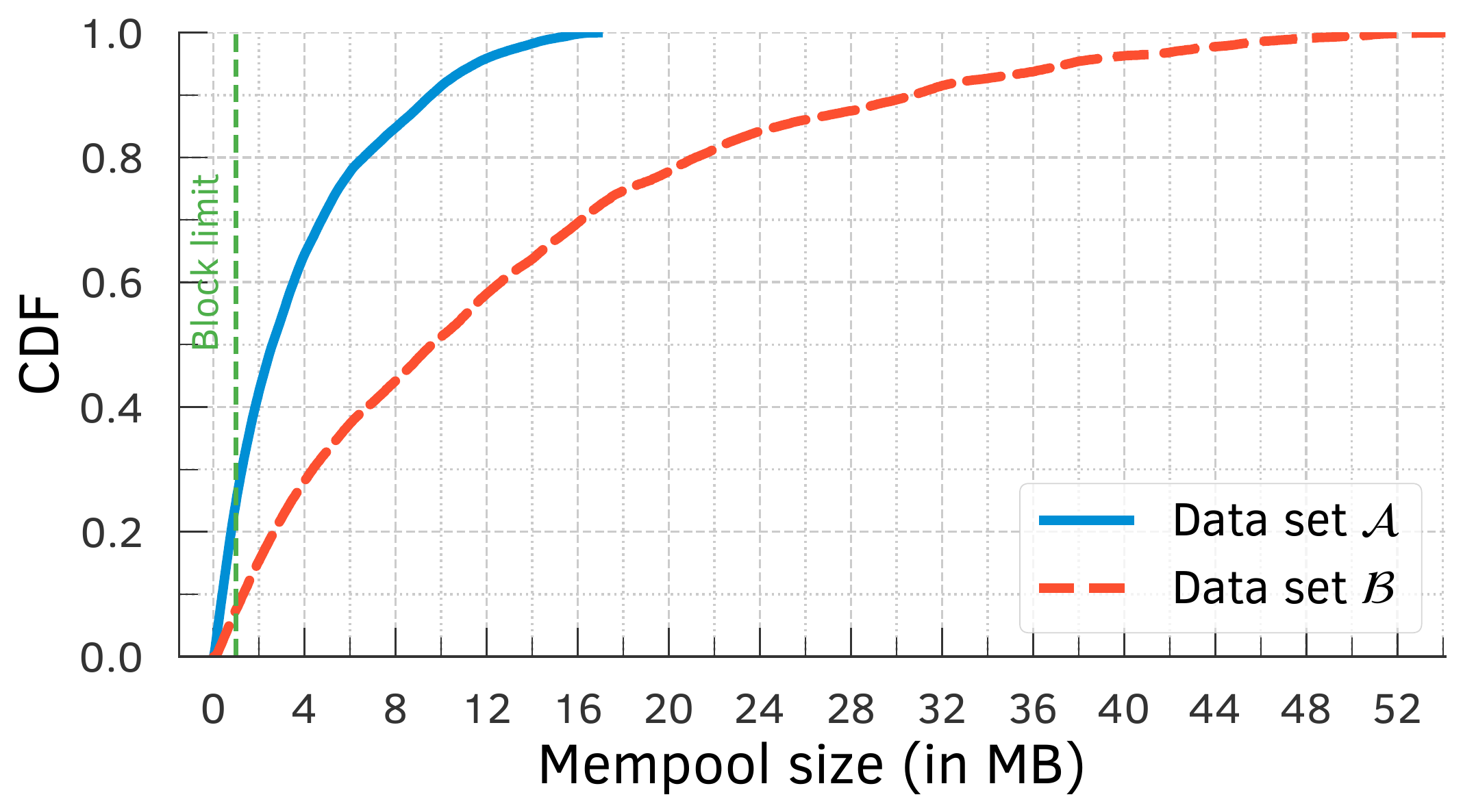}
        \sfigcap{}\label{fig:mempool-congestion}
    \end{subfigure}
    \begin{subfigure}[b]{\threecolgrid}
        \includegraphics[width={\textwidth}]{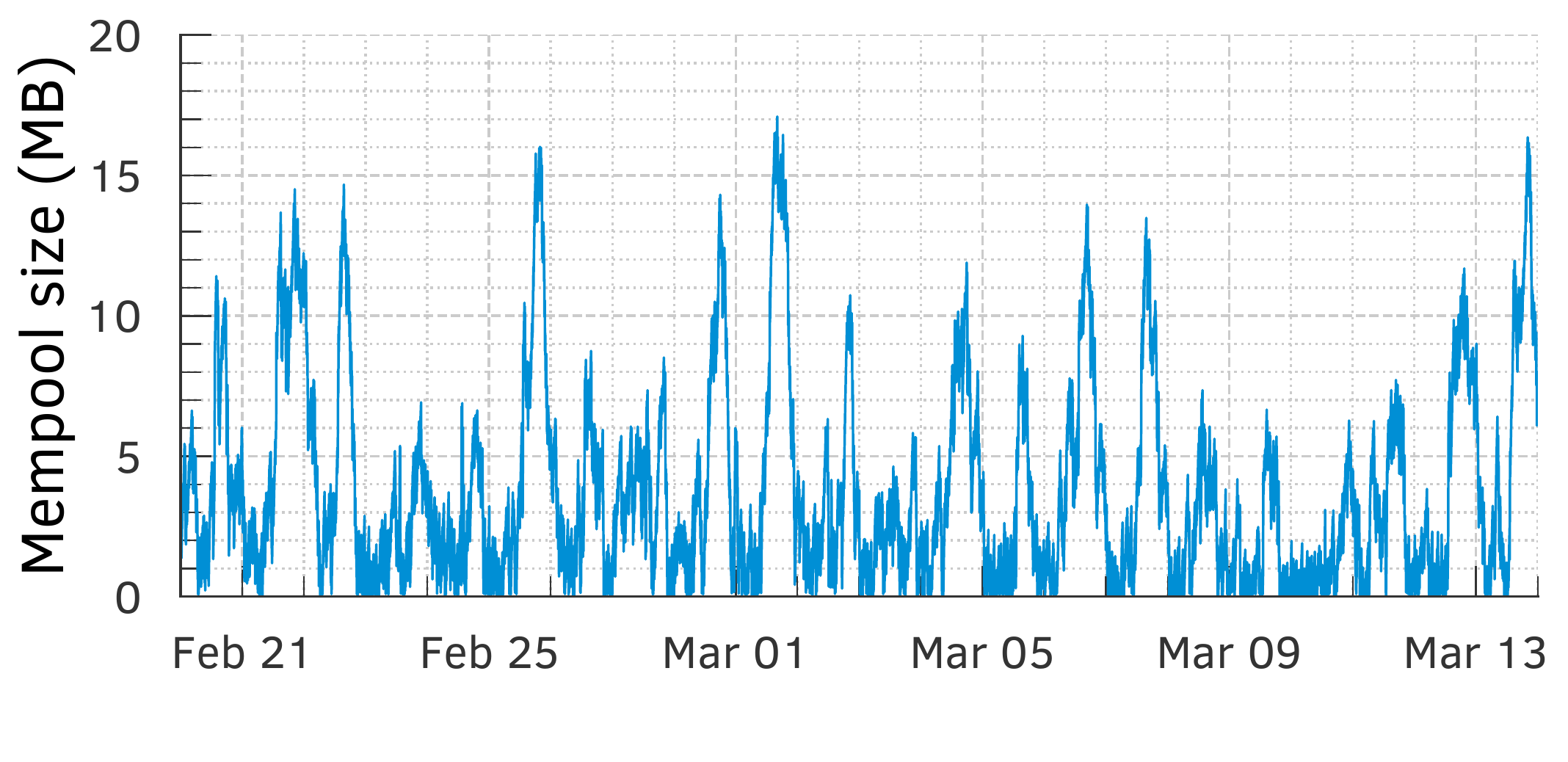}
        \sfigcap{}\label{fig:mpool-sz-a}
    \end{subfigure}
    \figcap{(a) Volume of transactions issued and blocks mined as a function of time, showing that transactions have been issued at high rates since mid-2017; (b) Distributions of \mpool{} size in both data sets \dsa{} and \dsb{}, and (c) the size \mpool in \dsa{} as a function of time, both indicating that congestion is typical in Bitcoin.}
\end{figure*}

\paraib{Data set \dsc{}.}
The insights derived from the above data motivated us to shed light on the
aberrant behavior of mining pool operators (MPOs).
To this end, we gathered all ($\num{53214}$) Bitcoin blocks mined and their \num{112542268} transactions from Jan. 1\tsup{st} to Dec. 31\tsup{st} 2020.
%
%
%
%
These blocks also contain one Coinbase transaction per block, which the MPO creates to receive the block and the fee rewards.
This data set, labeled \dsc{}, contains $\num{112489054}$ issued transactions (see Table~\ref{tab:datasets}).
MPOs typically include a \stress{signature} or \stress{marker} in the Coinbase
transaction, probably to claim their ownership of the block.
Following prior work (e.g., \cite{judmayer2017merged,Romiti2019ADD}), we use such
markers for identifying the MPO (owner) of each block.
We failed to identify the owners of $\num{703}$ blocks (or
approximately $1.32\%$ of the total), albeit we inferred $30$ MPOs in our data set.
In this paper, we consider only the top-20 MPOs whose combined normalized hash-rates account for $98.08\%$ of all blocks mined.
Figure~\ref{fig:dist-txs-blks-dataset-c} shows the count of blocks mined by the top-20 MPOs according to \dsc{}.
%
The top five MPOs in terms of the number of blocks ($B$) mined are F2Pool ($B$: $\num{9326}$; $h$: $17.53\%$), Poolin ($B$: $\num{7876}$; $h$: $14.80\%$), BTC.com ($B$: $\num{6381}$; $h$: $11.99\%$), 
 AntPool ($B$: $\num{5832}$; $h$: $10.96\%$), and Huobi ($B$: $\num{3990}$; $h$: $7.5\%$). 
We use this data set in \S\ref{sec:prioritization-norms} and \S\ref{sec:self-interest-scam-txs}.

\section{Analyzing Norm Adherence}
\label{sec:prioritization-norms}

In this section, we analyze whether Bitcoin miners adhere to prioritization norms, when selecting transactions for confirmation. To this end, we first investigate whether transaction ordering matters to Bitcoin users in practice, i.e., are there times when transactions suffer extreme delays and do users offer high transaction fees in such times to confirm their transactions faster? We then conduct a progressively deeper investigation of the norm violations, including potential underlying causes, which we investigate in greater detail in the subsequent sections.


\subsection{Does transaction ordering matter?} \label{sec:mempool}

\begin{figure*}[tbh]
    \centering
    \begin{subfigure}[b]{\threecolgrid}
        \includegraphics[width={\textwidth}]{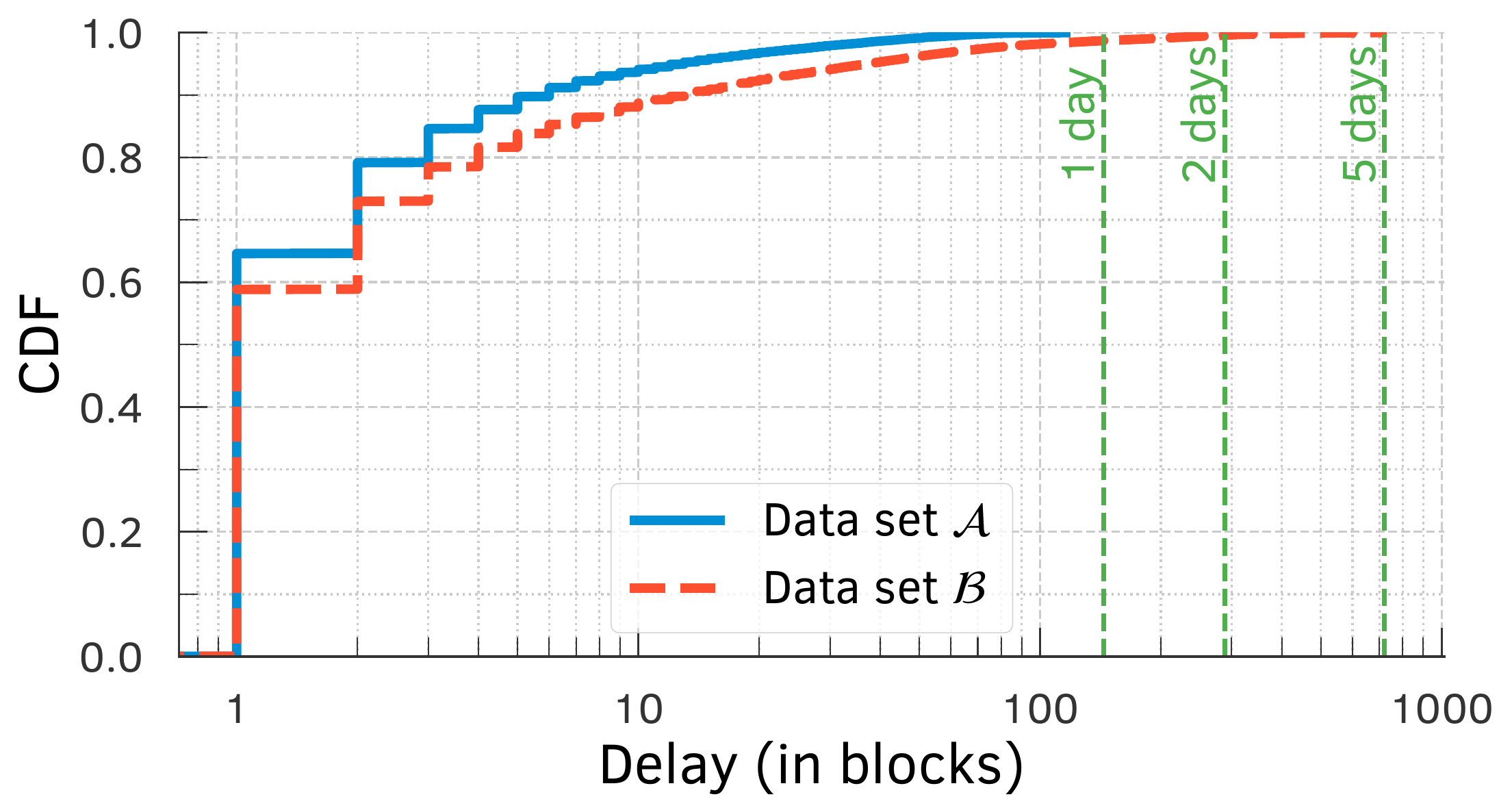}
        \sfigcap{}\label{fig:tx-commit-times}
    \end{subfigure}
    \begin{subfigure}[b]{\threecolgrid}
        \includegraphics[width={\textwidth}]{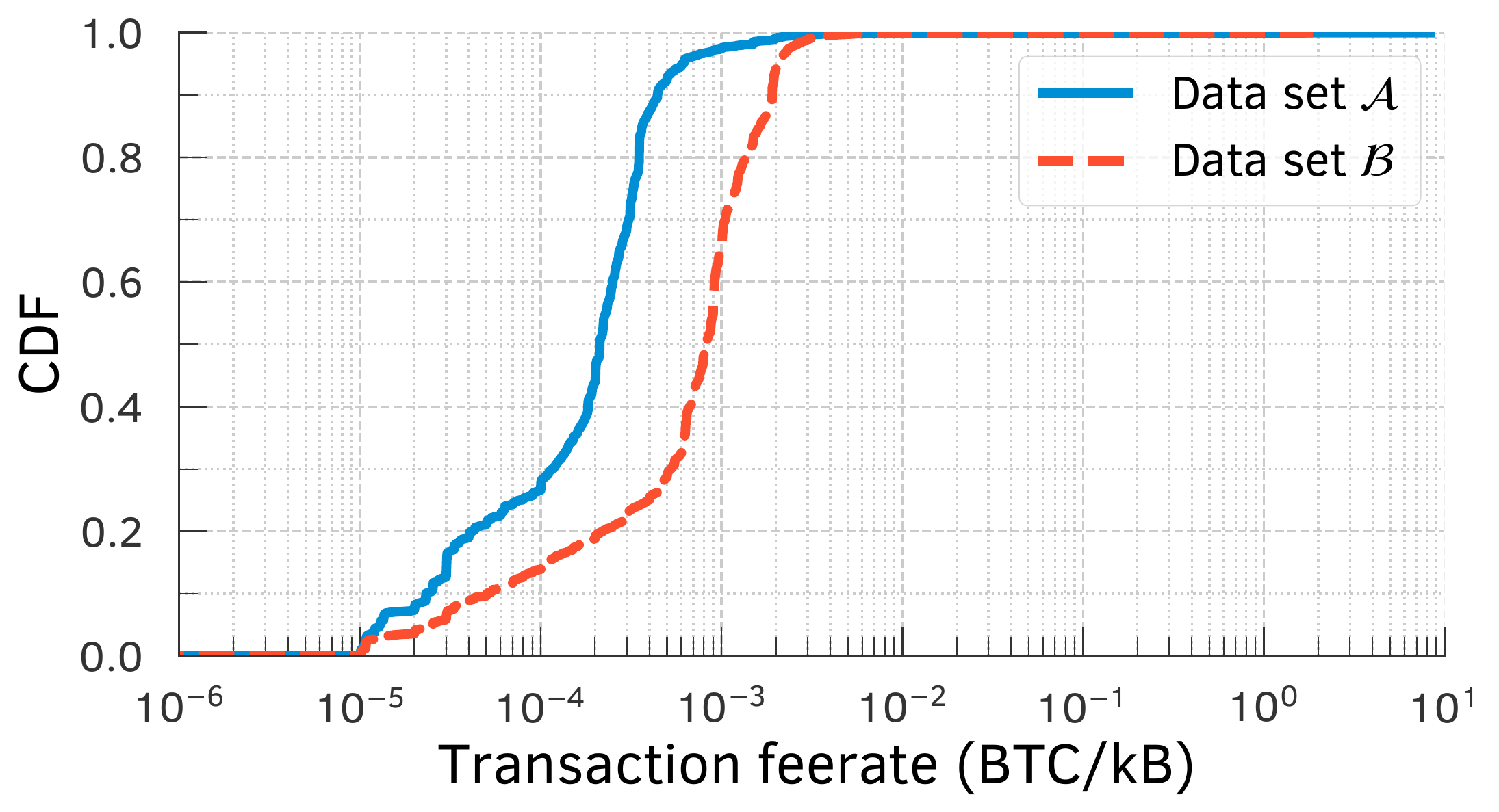}
        \sfigcap{}\label{fig:cdf-fee-all}
    \end{subfigure}
    \begin{subfigure}[b]{\threecolgrid}
        \includegraphics[width={\textwidth}]{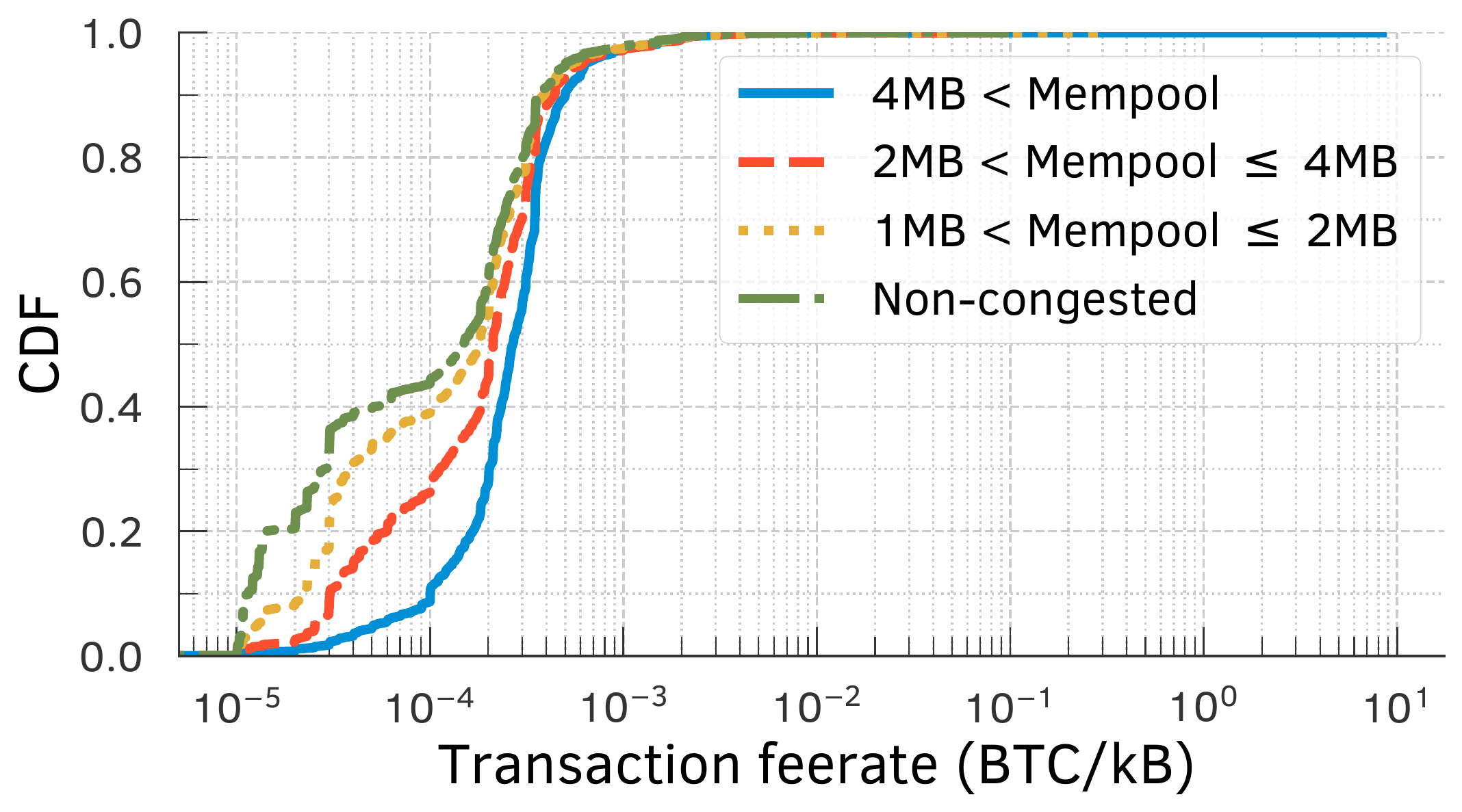}
        \sfigcap{}\label{fig:fee-cong-rel-a}
    \end{subfigure}
    \figcap{(a) Distributions of delays until transaction inclusion show that a significant fraction of transactions experience at least 3 blocks (or approximately 30 minutes) of delay; Distributions of fee-rates for (b) all transactions and (c) transactions (in \dsa{}) issued at different congestion levels clearly indicate that users incentivize miners through transaction fees.}
\end{figure*}



A~congestion in the \mpool leads to contention among transactions for
inclusion in a block.
Transactions that fail to contend with others (i.e., win a spot for inclusion)
experience inevitable delays in commit times.
Transaction ordering, hence, has crucial implications for users when the \mpool{} experiences congestion.
For instance, the Bitcoin Core code and most of the wallet software rely on the distribution of transactions' fee-rates included in previous blocks to suggest to users the fees that they should include in their transactions~\cite{BitcoinCore-2021,Fees@Coinbase,Lavi-WWW2019}.
Such transaction-fee predictions from any predictor, which assume that miners follow the norm, will be misleading.\footnote{%
Coinbase, one of the top cryptocurrency exchanges, does not allow users to set transaction fees manually. Instead it charges a fee based on how much they expect to pay for the concerned transaction, which in turn relies on miners following the norm~\cite{Fees@Coinbase}.}
%
%
Below, we examine whether \mpool{} in a real-world blockchain deployment
experiences congestion and its impact on transaction-commit
delays.
We then analyze whether, and how, users adjust transaction fees to cope with
congestion, and the effect of these fee adjustments on commit delays.

\subsubsection{Congestion and delays} \label{subsec:cong-delays}

Bitcoin's design---specifically, the adjustment of hashing difficulty to enforce
a constant mining rate---ensures that there is a steady flow of currency
generation in the network.
The aggregate number of blocks mined in Bitcoin, consequently, increases
linearly over time (Figure~\ref{fig:cdf-tx-blks-btc}).
Transactions, however, are \stress{not} subject to such constraints and have
been issued at much higher rates, particularly, according to Figure~\ref{fig:cdf-tx-blks-btc}, since mid-2017:
$60\%$ of all transactions ever introduced were added in only in the last 3.5 years of the nearly decade-long life of the cryptocurrency.
Should this growth in transaction issues continue to hold, transactions will
increasingly have to contend with one another for inclusion within the limited
space (of $\uMB{1}$) in a block.
Below, we empirically show that this contention among transactions is
already common in the Bitcoin network.

Using the data sets \dsa{} and \dsb{} (refer~\S\ref{sec:datasets}), we measured
the number of unconfirmed transactions in the \mpool{}, at the granularity of
\usd{15}.
Per Figure~\ref{fig:mempool-congestion}, congestion in \mpool{} is
\stress{typical} in Bitcoin:
During the three-week period of \dsa, the aggregate size of all unconfirmed
transactions was above the maximum block size (of $\uMB{1}$) for nearly $75\%$
of the time; per data set \dsb the \mpool was congested for nearly $92\%$ of the
time period.
%
%
%
Figure~\ref{fig:mpool-sz-a} provides a complementary view of the \mpool
congestion in \dsa{}, by plotting the \mpool{} size as a function of time.
The measurements reveal a huge variance in \mpool{} congestion, with size of
unconfirmed transactions at times exceeding 15-times the maximum size of a
block.
Transactions queued up during such periods of high congestion will have to
contend with one another until the \mpool{} size drains below $\uMB{1}$.
These observations also hold in data set \dsb{}, the details of which are in~\S\ref{sec:supp-tx-ord}.

The \mpool congestion, which in turns leads to the contention among transactions
for inclusion in a block, has one serious implication for users: delays in
transaction-commit times.
While $65\%$ ($60\%$) of all transactions in data set \dsa (\dsb) get committed
in the next block (i.e., in the block immediately following their arrival in the
\mpool), Figure~\ref{fig:tx-commit-times} shows that nearly $15\%$ ($20\%$) of
them wait for at least $3$ blocks (i.e., $30$~minutes on average).
Moreover, $5\%$ ($10\%$) of the transactions wait for $10$ or more blocks, or
$100$~minutes on average, in data set \dsa (\dsb).
While no transaction waited for more than a day in data set \dsa, a~small
percentage of transactions waited for up to five days (because of the high
levels of congestion in June 2019) in data set \dsb.

\parai{Takeaways.}
\mpool{} is typically congested in Bitcoin. Transactions, hence, typically
contend with one another for inclusion in a block.
The \mpool{} congestion has non-trivial implications for transaction-commit times.

\subsubsection{Transaction fee-rates and delays}\label{subsec:feerates}

To combat the delays and ensure that a transaction is committed ``on time''
(i.e., selected for inclusion in the earliest block), users may include a
transaction fee for incentivizing the miner.
While the block reward from May $11$, 2020 is
$\uBTC{6.25}$, the aggregate fees accrued per block is becoming considerable
(i.e., $6.29\%$ of the total miner revenue in 2020 per Table~\ref{tab:fee-revenue} in~\S\ref{sec:signif-tx-fees}).
%
%
%
Prior work also show that revenue from transaction fees is
clearly increasing~\cite{Easley-SSRN2017}.
With the volume of transactions growing aggressively (Figure~\ref{fig:cdf-tx-blks-btc}) over time and the block rewards, in Bitcoin,
halving every four years, it is inevitable that transaction fees will be an
important, if not the only, criterion for including a transaction.
Below, we analyze whether Bitcoin users incentivize miners via transaction fees and if such incentives are effective today.


Per Figure~\ref{fig:cdf-fee-all} the transaction fee-rate of committed
transactions in both data sets \dsa{} and \dsb{} exhibits a wide range, from
$10^{-6}$ to beyond $\uTxFee{1}$.
The fee-rate distributions of committed transactions also do not vary much between different mining pool operators (refer Figure~\ref{fig:cdf-fee-top5} in \S\ref{sec:tx-fees-across-mpos}).
A~few transactions ($0.001\%$ in \dsa{} and $0.07\%$ in \dsb{}) were committed,
despite offering fee-rates less than the recommended minimum of
$10^{-5}$~\feeunit{}.
A non-trivial percentage of transactions offered fee-rates that are two orders
of magnitude higher than the recommended value; particularly, in data set \dsb,
perhaps due to the comparatively high levels of congestion (cf.
Figure~\ref{fig:mpool-sz-a} and Figure~\ref{fig:mpool-sz-b}), $34.7\%$ of transactions offered fee-rates higher
than $10^{-3}$~\feeunit{}.
Approximately $70\%$ ($51.3\%$) of the transactions in data set \dsa{} (\dsb{})
offer fee-rates between $10^{-4}$ and $10^{-3}$~\feeunit{}, i.e., between one
and two orders of magnitude more than the recommended minimum.
Such high fee-rates clearly capture the users’ intents to incentivize the
miners.

Our premise is that the (high) fee-rates correlate with the level of
\mpool{} congestion.
Said differently, we hypothesize that users increase the fee-rates to curb the
delays induced by congestion.
To test this hypothesis, we separate the \mpool{} snapshots (cf.~\S\ref{subsec:cong-delays}) into $4$ different bins.
Each bin corresponds to a specific level of congestion identified by the \mpool{} size as follows:
lower than \uMB{1} (\stress{no congestion}), in $(1, 2]$ MB (\stress{lowest congestion}), in $(2,
4]$ MB, and higher than \uMB{4} (\stress{highest congestion}).
The fee-rates of transactions observed in the different bins or congestion levels, in Figure~\ref{fig:fee-cong-rel-a}, then validates our hypothesis:
Fee-rates are strictly higher (in distribution, and hence also on average) for higher congestion levels.

\begin{figure}[tb]
    \centering
    \includegraphics[width={\onecolgrid}]{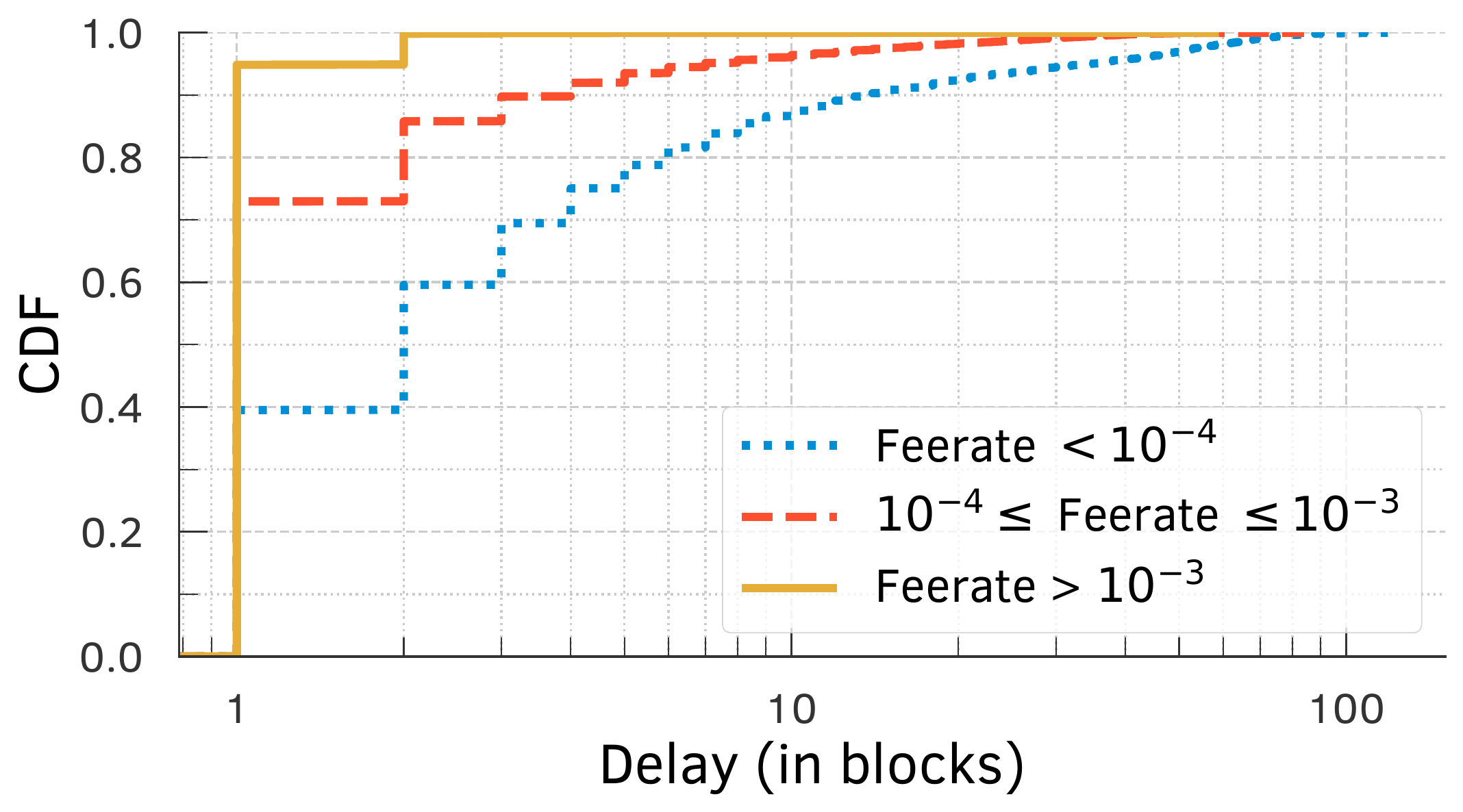}
    \figcap{Distributions of transaction-commit delays for different fee-rates for transactions in \dsa{}; incentivizing miners via fee-rates works well in practice.}\label{fig:fee-delay-rel-a}
\end{figure}

Figure~\ref{fig:fee-delay-rel-a} shows that users' strategy of increasing
fee-rates to combat congestion seems to work well in practice.
Here, we compare the CDF of commit delays of transactions with low (i.e., less
than $10^{-4}$~\feeunit{}), high (i.e., between $10^{-4}$ and
$10^{-3}$~\feeunit{}), and exorbitant (i.e., more than $10^{-3}$) fee-rates, in data set \dsa{}.
Similar analysis with data set \dsb{} is provided in~\S\ref{sec:fees-and-cong}.
We observe that an increase in the transaction fee-rates is consistently
rewarded (by miners) with a decrease in the commit delays.
This observation suggests that, at least to some extent, miners prioritize
transactions for inclusion based fee-rates or the fee-per-byte metric.

\parai{Takeaways.}
A significant fraction of transactions offer fee-rates that are well above the
recommended minimum.
Fee-rates are typically higher at higher congestion levels, and reduce the
commit delays.
These observations suggest that users are indeed willing to spend money to
decrease commit delays for their transactions during periods of congestion.

\subsection{Do miners follow the norms?}\label{subsec:mining-prioritization-based-feerate}

Whether miners follow the transaction prioritization norms (as widely assumed)
has implications for both Bitcoin and its users:
The software used by users, for instance, assumes an adherence to these norms
when suggesting a transaction fee to the user~\cite{BitcoinCore-2021,Fees@Coinbase,Lavi-WWW2019}.
Deviations from these norms, hence, have far reaching implications for both the
blockchain and crucially for Bitcoin users.


\subsubsection{Fee-rate based selection when mining new blocks}

Our finding above that transactions offering higher fee-rates experience lower confirmation
delays suggests that miners tend to account for transaction fee-rates when choosing transactions for new blocks. 
We now want to check, however, if transaction fee-rate is the primary or the sole determining factor in transaction selection.
To this end, we check our data sets for transaction pairs, where one transaction was issued earlier and has a higher fee-rate than the other, but was committed later than the other.
The existence of such transaction pairs would unequivocally show that fee-rate alone does not explain the order in which they are selected.

We sampled $30$ \mpool snapshots, uniformly at random, from the set of all available snapshots in data set \dsa{}.
Suppose that, in each snapshot, we denote, for any transaction $i$, the time at
which it was received in the \mpool by $t_i$, its fee-rate by $f_i$, and the
block in which it was committed by $b_i$.
We then selected, from each snapshot, all pairs of transactions $(i,j)$ such
that $t_{i} < t_{j}$ and $f_{i} > f_{j}$, but $b_{i} > b_{j}$.
Such pairs clearly constitute a violation of the fee-rate-based transaction-selection norm.

\begin{figure}[tb]
   \centering
   \begin{subfigure}[b]{\onecolgrid}
     \includegraphics[width={\textwidth}]{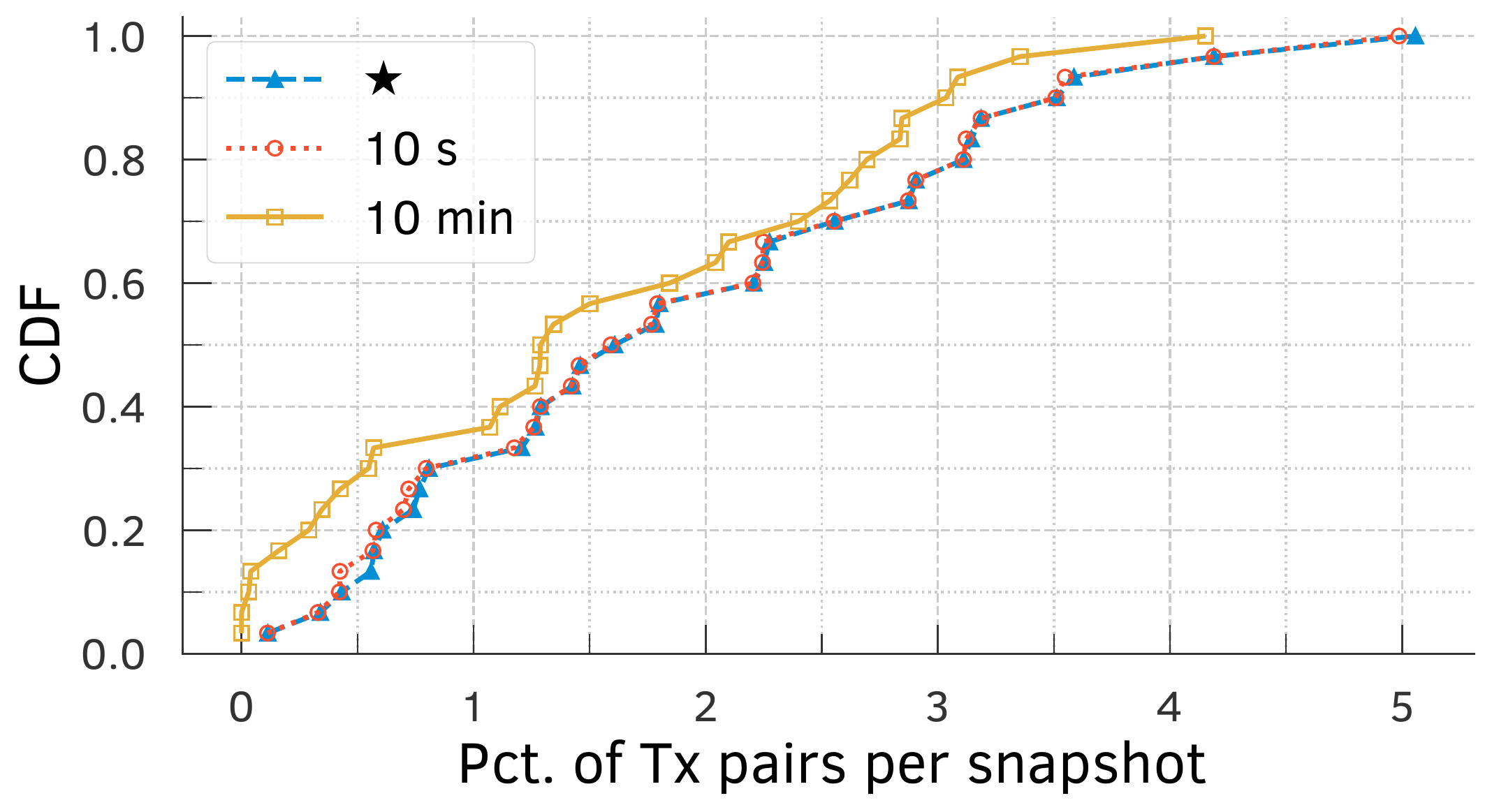}
     \sfigcap{All transactions}		
     \label{fig:violation-all}		
   \end{subfigure}		
   \begin{subfigure}[b]{\onecolgrid}		
     \includegraphics[width={\textwidth}]{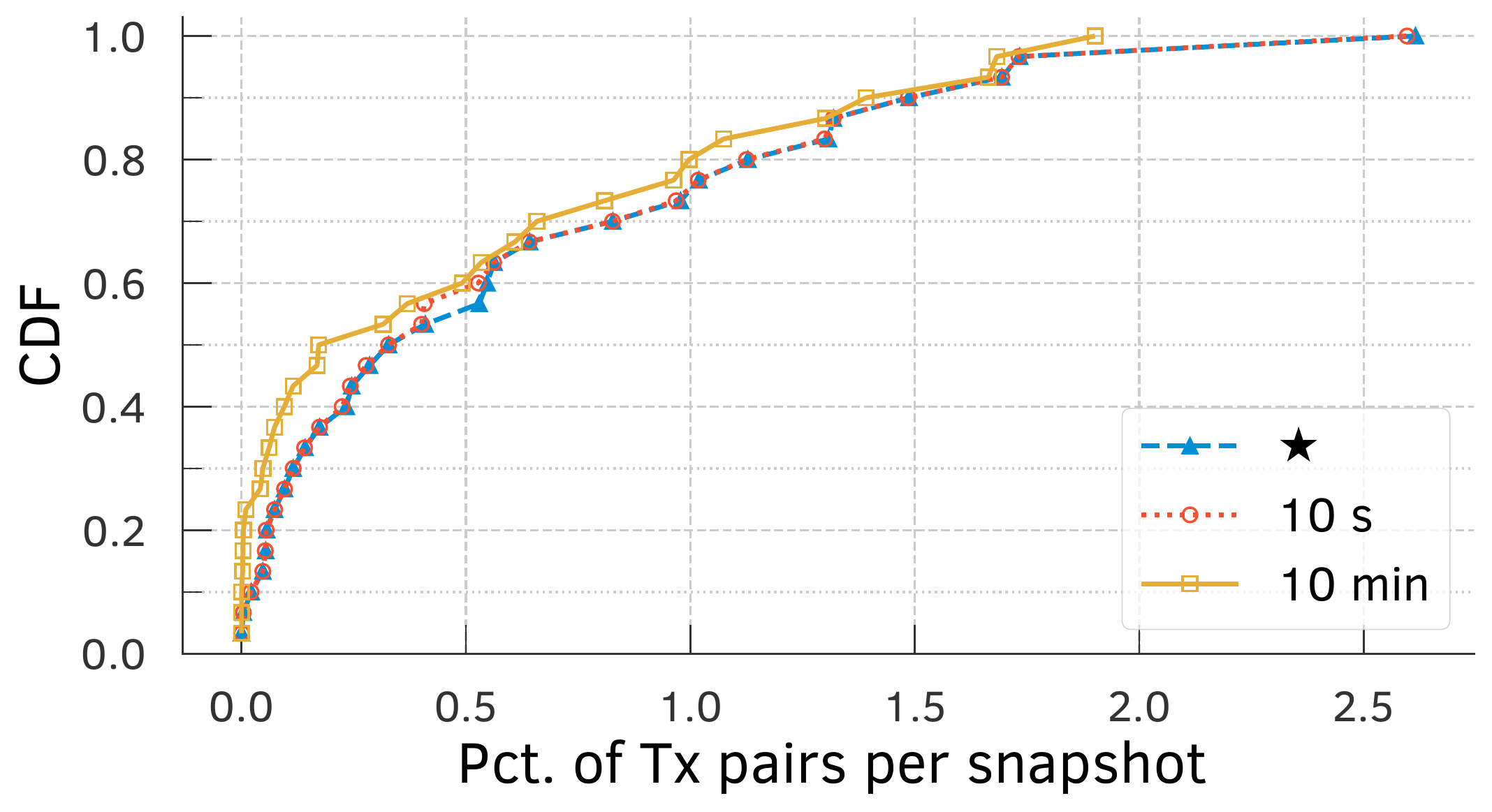}		
     \sfigcap{Only Non-CPFP transactions}		
     \label{fig:violation-non-cpfp}		
   \end{subfigure}		
   \figcap{There exists a non-trivial fraction of transaction pairs violating the norm across all		
     snapshots, clearly indicating that miners do \underline{not} adhere to the norm.}
   \label{fig:violation}
\end{figure}

Figure~\ref{fig:violation-all} shows a cumulative distribution of the fraction of all transaction pairs (line labeled ``$\star$'') violating the norm over all sampled
snapshots. 
Across all snapshots, a small but non-trivial fraction of all transaction pairs violate the norm. 
One potential explanation for violations might be that the transactions are received by the mining pools in different order than the one in which our \mpool receives.
To account for such differences, we tighten the time constraint as $t_{i} + \epsilon < t_{j}$ and use an $\epsilon$ of either $10$ seconds or $10$~minutes. 
Even with the tightened time constraints, Figure~\ref{fig:violation-all} shows that a non-trivial fraction of 
all transaction pairs violate the norm.

Another potential source of violations are Bitcoin's dependent (or, parent and child) transactions, where the child pays a high fee to incentivize miners to also confirm the
parent from which it draws its inputs. This mechanism enables users to
``accelerate'' a transaction that has been ``stuck'' because of low
fee~\cite{CoinStaker-2018}. As the existence of such \newterm{child-pays-for-parent (CPFP)} transactions (formally defined in~\S\ref{sec:cpfp-txs})
would introduce false positives in our analysis we decided to discard them.
Figure~\ref{fig:violation-non-cpfp} shows that the violations exist even after discarding all such dependent transaction pairs.

\subsubsection{Fee-rate based ordering within blocks}

We now turn our attention to transaction ordering within individual (mined) blocks in Bitcoin.
If a miner followed GBT, transactions would be ordered based on their fee-rate.
In this case, given the set of non-CPFP transactions $T = \{T_1, T_2, .... T_n\}$ included in a block $B$, we should be able to predict their position in the block by simply ordering the transactions based on their fee-rate (as specified in the GBT implementation in Bitcoin Core).
To quantify the deviation from the norm, we compute a measure that we call \textit{\textbf{position prediction error (PPE)}}: PPE of a block $B$ is the average absolute difference between the predicted and the observed (actual) positions for all transactions in block $B$, normalized by the size of the block ($n$) and expressed as a percentage. More precisely,

\begin{align*}
    PPE(B) = \sum_{i=1}^n \dfrac{(|T^{p}_{i} -  T^{o}_{i})|) \cdot 100}{n}
\end{align*}

where $T^{p}_{i}$ and $T^{o}_{i}$ are the predicted and observed positions of a transaction, respectively.

\begin{figure}[tb]
  \centering
  \begin{subfigure}[b]{\onecolgrid}
    \includegraphics[width={\textwidth}]{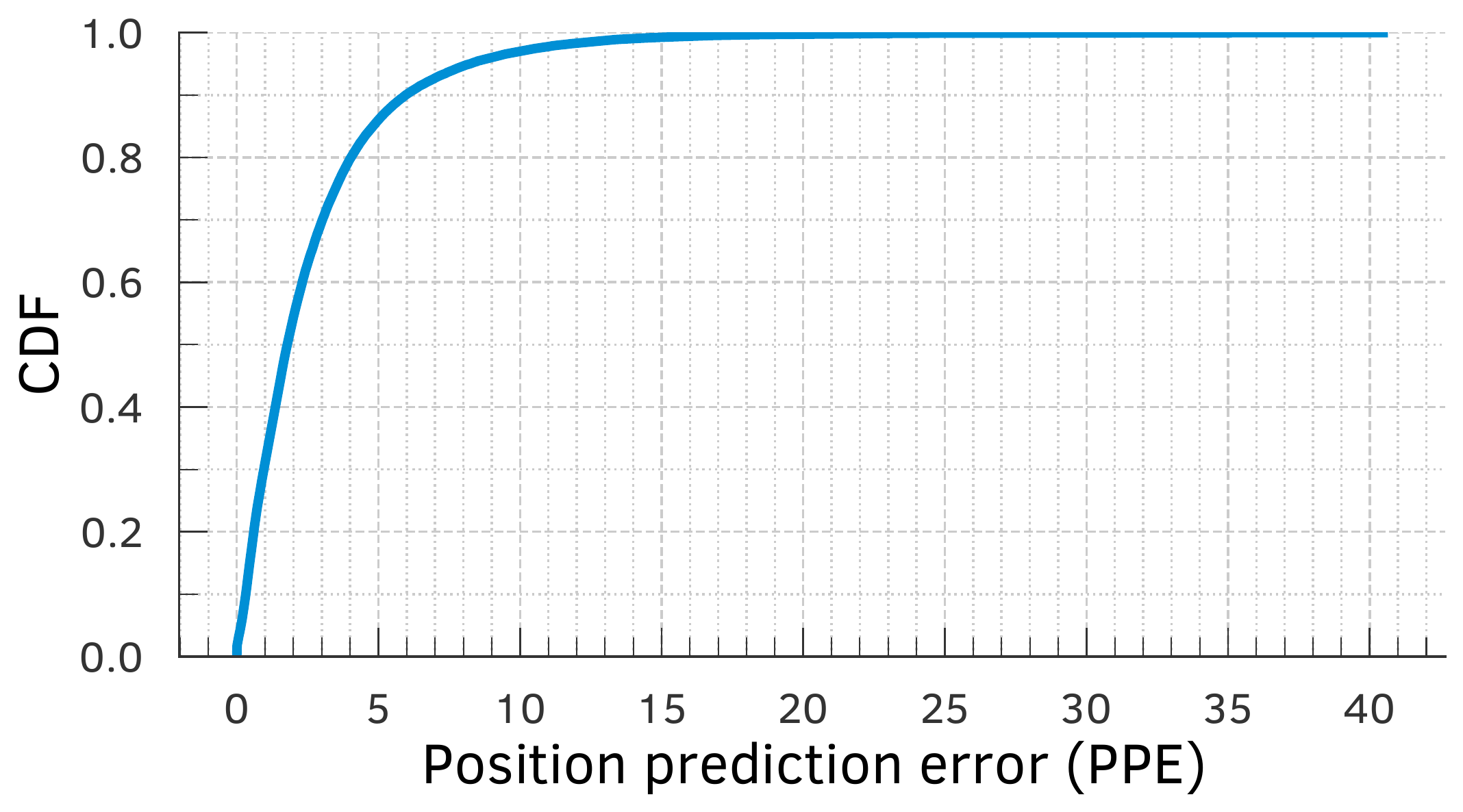}
    \sfigcap{Overall position prediction error}\label{fig:deviation-within-blocks-overall}
  \end{subfigure}
  \begin{subfigure}[b]{\onecolgrid}
    \includegraphics[width={\textwidth}]{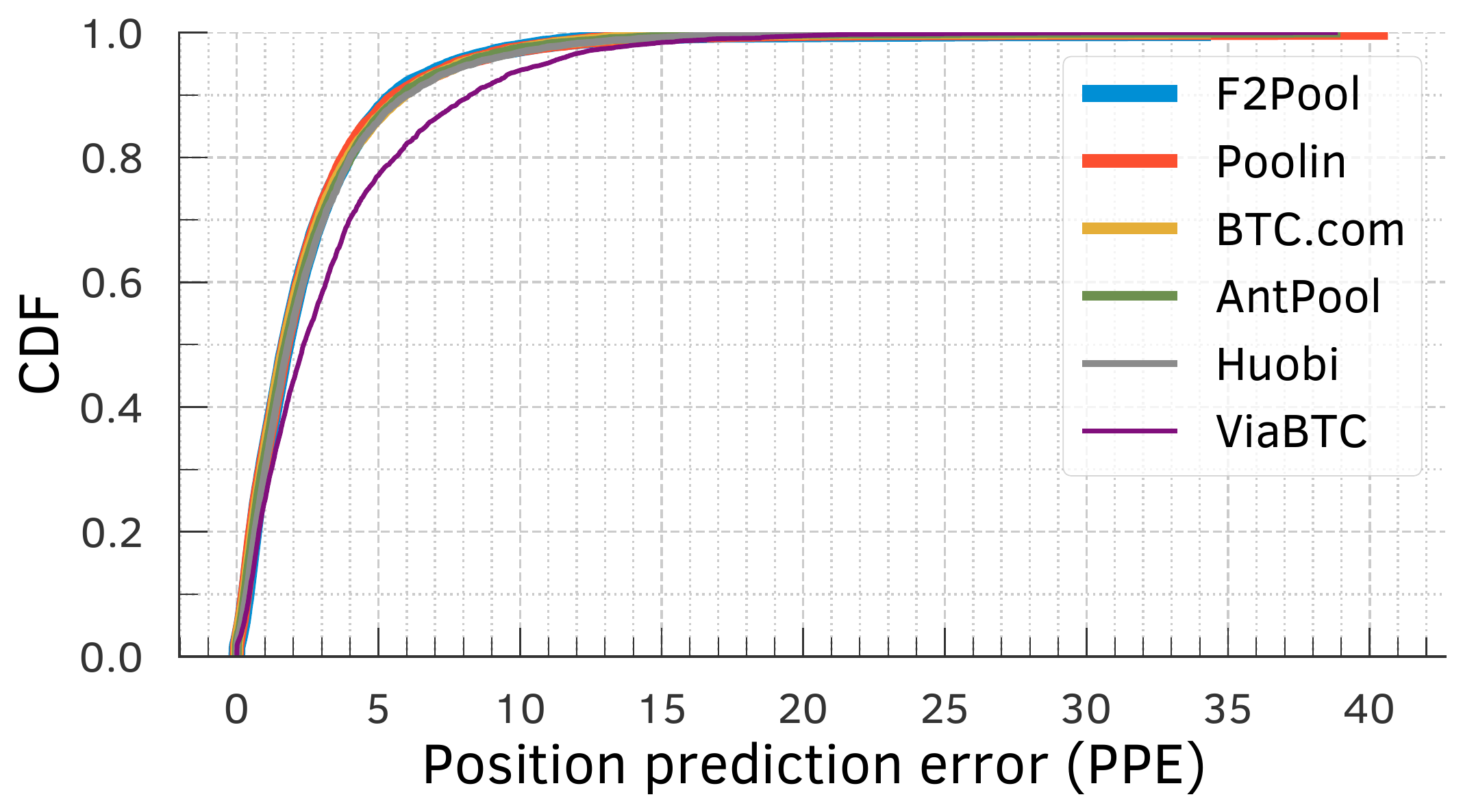}
    \sfigcap{Position prediction error of the top-6 MPOs}\label{fig:deviation-within-blocks-top6-mpo}
  \end{subfigure}
  \figcap{
  Position prediction error (PPE). (a) There are 52,974 (99.55\%) blocks with at least one non-CPFP txs. The mean PPE is 2.65\%, with an std of 2.89. 80\% of all blocks has PPE less than 4.03\%. (b) The PPEs of blocks mined by the top-6 MPOs according to their normalized hash rate.}
  \label{fig:deviation-within-blocks}
\end{figure}

Figure~\ref{fig:deviation-within-blocks-overall} shows the cumulative distribution of PPE values for each block in our data set \dsc{}, containing \num{53214} blocks. $80\%$ of the blocks have PPE values less than $4.03\%$. The mean PPE across all blocks is $2.65\%$, with a standard deviation of $2.89$.
Per this plot the position of a transaction within a block can be predicted with very high accuracy (within a few percentile position error), suggesting that transactions are by and large ordered within a block based on their fee-rate. Figure~\ref{fig:deviation-within-blocks-top6-mpo} shows PPE values separately for each of the $6$ largest mining pools in data set \dsc{}. The plots show that all mining pools by and large follow the norm, though some like ViaBTC seems to deviate slightly more from the norm compared to the other mining pools.

\subsubsection{Fee-rate threshold for excluding transactions}


In their default configuration, many nodes in the Bitcoin P2P network drop (i.e., ignore) transactions that offer less than a threshold fee-rate (typically, $10^{-5}$ \feeunit). 
As miners select transactions for inclusion from their local Bitcoin P2P node, this (default) norm would result in such low-fee transactions never being included in the blockchain, even during periods of non-congestion (when blocks have spare capacity to accommodate additional transactions).

We collected data set \dsa{} using a default Bitcoin node, and our node, hence, did not accept or record low-fee transactions.
When gathering data set \dsb{}, however, we configured our Bitcoin node to accept all transactions, irrespective of their fee-rates.
In data set \dsb{}, our node, consequently, received \num{1084} transactions that offered less than the recommended fee-rate and $489$ ($45.11\%$) of them were zero-fee transactions.
From these low fee-rate transactions, only \num{53} ($4.89\%$) were confirmed in the Bitcoin blockchain; $9$ ($16.98\%$) were confirmed months after they were observed in our data set.
In contrast, the vast majority ($99.7\%$) of the transactions that offered greater than or equal to the recommended fee-rate were all (eventually) confirmed.
%
%
Interestingly, the low-fee transactions were confirmed by just three mining pools: F2Pool, ViaBTC, and BTC.com included $38$, $14$, and $1$ low-fee transactions, respectively.
Our findings suggest that while the norm of ignoring transactions offering less than the recommended fee-rate is being by and large followed by all miners, a few occasionally deviate from the norm.

\section{Investigating Norm Violations} \label{sec:self-interest-scam-txs}

Our analysis so far showed that while Bitcoin miners by and large follow transaction-prioritization norms, there are many clear instances of norm violations. 
Our next goal is to develop a deeper understanding of the underlying reasons or motivations for miners to deviate from the fee-rate based norms, at least for some subset of all transactions. 
To this end, we focus our investigation on the following three types of transactions, where we hypothesize miners might have an incentive to deviate from the current norms, which are well-aligned towards maximizing their rewards for mining.

\begin{enumerate}
  \item {\it Self-interest Transactions:} Miners have a vested interest in a transaction, where the miners themselves are a party to the transaction, i.e., a sender or a receiver of bitcoins. Miners may have an incentive to selfishly accelerate the commitment of such transactions in the blocks mined by themselves. 
  
  \item {\it Scam-payment Transactions:} Bitcoins are increasingly being used to launch a variety of ransomware and scam attacks~\cite{Frenkel@nyt17,Mathews@forbest17,Frenkel@nyt20}. A recent scam attack involved using hijacked Twitter accounts of celebrities to encourage their followers to send bitcoins to a specific Bitcoin wallet address~\cite{Frenkel@nyt20}.
  Given the timely and widespread coverage of this attack in popular press and other similar attacks on crowd-sourced websites for reporting scam transactions~\cite{Scam@BitcoinAbuse,Scam@ScamAlert}, and with governments trying to blacklist wallet addresses of entities suspected of illegal activities~\cite{De@Coindesk,Hinkes@Coindesk}, we hypothesize that some miners might decelerate or even absolutely exclude the commitment of scam-payment transactions out of fear or ethical concerns.
  
  \item {\it Dark-fee Transactions:} Recently, some mining pool operators have started offering transaction acceleration services~\cite{BTC@accelerator,ViaBTC@accelerator,Poolin@accelerator,F2Pool@accelerator,AntPool@accelerator}, where anyone wanting to prioritize their transactions can pay an additional fee to a specific mining pool via a side-channel (often, the MPO's website or via a private-channel~\cite{strehle2020exclusive}). Such transaction fees are ``dark'' or opaque to other mining pools and the public, and we hypothesize that some of the committed low-fee transactions might have been accelerated by using such services. 
  
\end{enumerate}

To detect whether a mining pool has accelerated or decelerated the above types of transactions, we first design a robust statistical test.
Later, we report our findings from applying the test on the three types of transactions.

\subsection{Statistical test for differential prioritization}

Our goal here is to propose a robust statistical test for detecting whether a given mining pool $m$ is prioritizing a given set of committed transactions $c$ \stress{differently} than all other miners.
The basic idea behind the statistical test is as follows. Suppose a mining pool is accelerating (decelerating) transactions in set $c$. In that case, these transactions will have a disproportionately high (low) chance of being included in blocks mined by this mining pool compared to the mining pool's hashing power (or rate).


\subsubsection{Test for differential transaction acceleration}

Consider a miner $m$ with normalized hash rate $h = \theta_0$ (estimated as fraction of blocks mined by $m$). Assume that we are given a set of transactions, denoted as $c$-transactions (for committed transactions), for which we wish to test whether miner $m$ is treating them preferentially. 

To test whether $m$ is prioritizing $c$-transactions, we look at all blocks that include at least one $c$-transaction, call them $c$-blocks. Suppose that there are $y$ such blocks.
If $m$ is not prioritizing $c$-transactions, then a fraction $\theta_0$ of all $c$-blocks should be $m$-blocks (i.e., mined by $m$); if $m$ is prioritizing $c$-transactions (compared to other miners) then the fraction will be higher. We want to test whether the true fraction $\theta$ is indeed $\theta_0$ or is higher. We formalize this as follows: We assume that each $c$-block has a probability $\theta$ to be an $m$-block and do the following test.
\begin{align*}
  & H_0: \theta = \theta_0 \\ 
  & H_1: \theta > \theta_0.
\end{align*}
Assuming that the observed number of $c$-blocks that are mined by $m$ is $x$, the $p$-value of the test is 
\begin{align*}
  p = Pr (B \ge x), 
\end{align*}
where $B$ is a binomial distribution of parameter $\theta_0$ and $y$, that is 
\begin{align*}
  p = \sum_{k=x}^y \binom{y}{k} \theta_0^k (1-\theta_0)^{(y-k)}.
\end{align*}
We may fix the size of the test (i.e., the maximal probability of type I error that corresponds to rejecting $H_0$ when $H_0$ is true) to $\alpha = 0.01$. Then $H_0$ should be rejected whenever $p<\alpha$. The smaller $p$, the higher the confidence in rejecting $H_0$, that is declaring that $m$ prioritizes c-transactions. 
%


The above test is relative in the sense that we can only detect if a miner treats $c$-transactions more preferentially than the rest of the miners. This test cannot conclude on whether it is the miner accelerating the $c$-transactions (relative to their deserved, i.e., fee-rate based, priority) or the rest of the miners are decelerating them. 
So, we look at additional empirical evidence from the position of the $c$-transactions within the $c$-blocks that include them.
Specifically, given the set of $c$-transactions $\{c_1, c_2, .... c_n\}$ committed by a miner $m$, 
%
we compute a measure that we call \textit{\textbf{signed position prediction error (SPPE)}} as the average signed difference between the predicted and observed  positions (measured as percentile rank) for all $c$-transactions within the blocks committed by $m$. More precisely,

\begin{align*}
  SPPE(m) = \dfrac{\sum_{i=1}^n (c^{p}_{i} -  c^{o}_{i}) \cdot 100}{n}
\end{align*}
where $c^{p}_{i}$ and $c^{o}_{i}$ are the predicted and the observed (percentile rank) positions, respectively, of transaction $c_i$ within the blocks committed by $m$.

\subsubsection{Test for differential transaction deceleration}

While the previous test checks for prioritization (or acceleration), one may also want to test for deceleration. To that end, a symmetric test can be used. Specifically, with the previous notation, the test would be
\begin{align*}
  & H_0: \theta = \theta_0 \\ 
  & H_1: \theta < \theta_0;
\end{align*}
and its $p$-value would be 
\begin{align*}
  p = Pr (B \le x), 
\end{align*}
where $B$ is a binomial distribution of parameter $\theta_0$ and $y$, that is 
\begin{align*}
  p = \sum_{k=0}^x \binom{y}{k} \theta_0^k (1-\theta_0)^{(y-k)}.
\end{align*}

\begin{figure}[b]
    \centering
    \begin{subfigure}[b]{\onecolgrid}
        \includegraphics[width={\textwidth}]{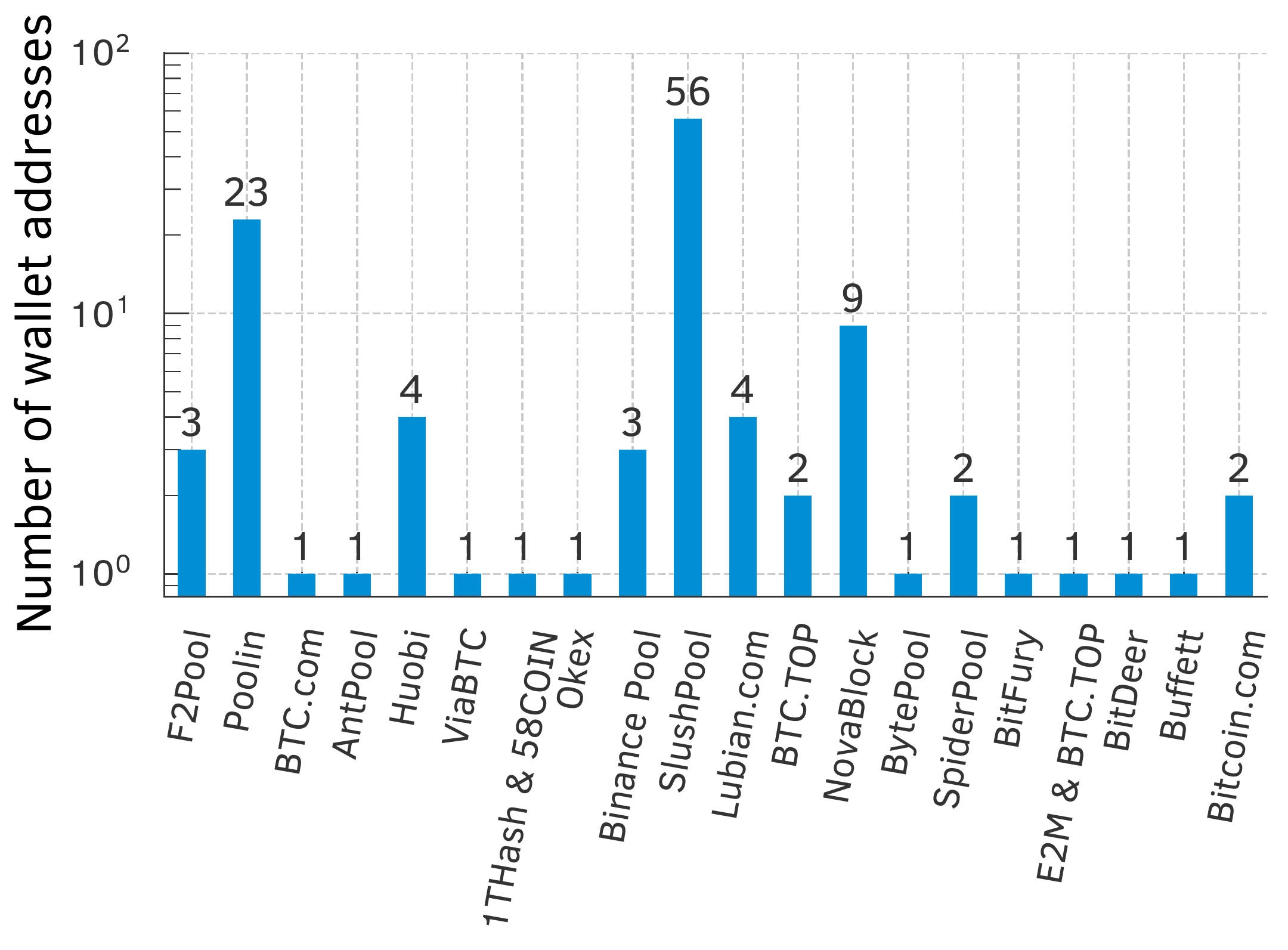}
        \sfigcap{}\label{fig:bar-number-wallet-addresses}
    \end{subfigure}
    \begin{subfigure}[b]{\onecolgrid}
        \includegraphics[width={\textwidth}]{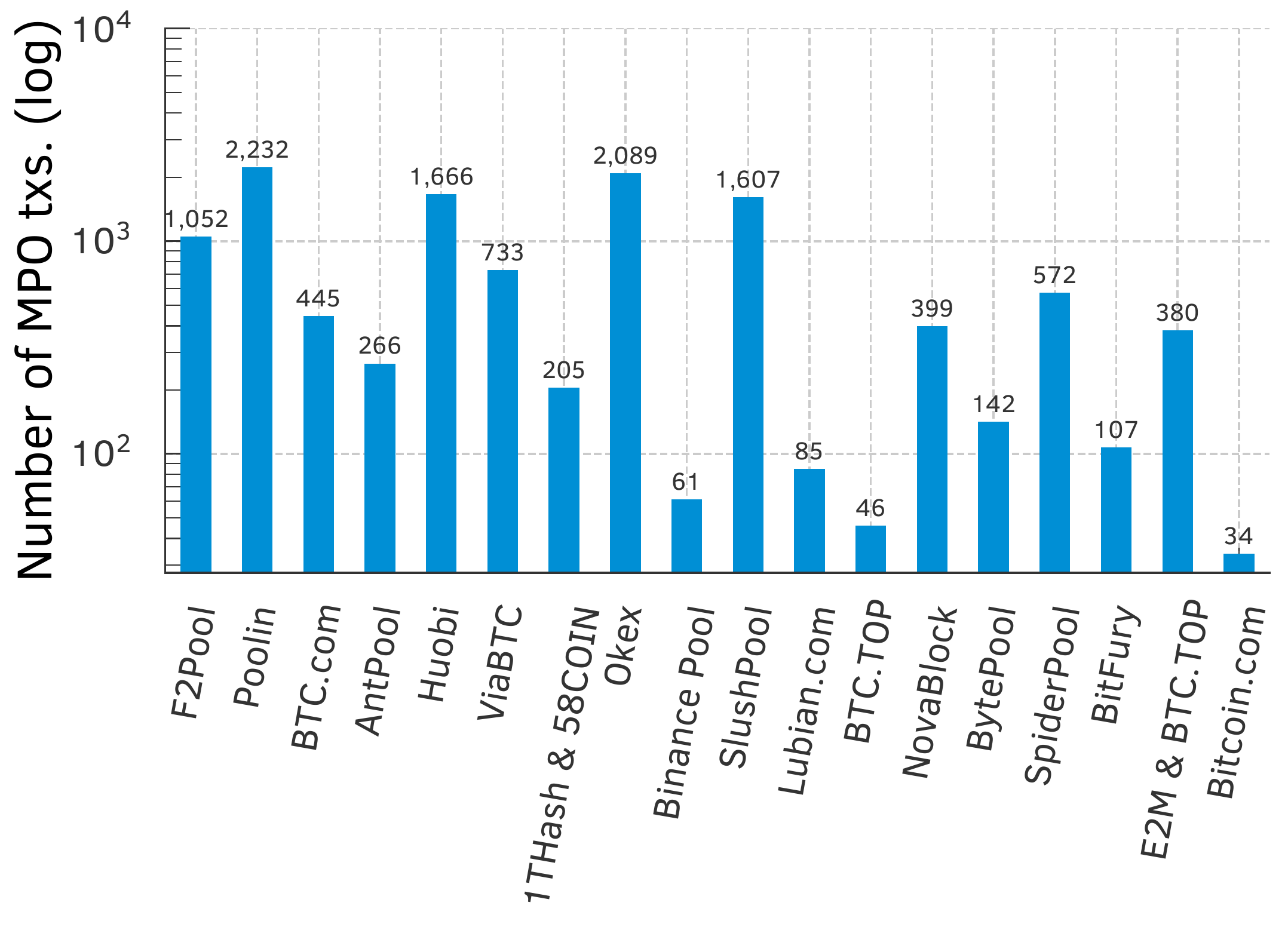}
        \sfigcap{}\label{fig:bar-num-wallet-addresses-mpo}
    \end{subfigure}
    \figcap{
    (a) Distribution of the number of wallet addresses used by each of the top-20 MPOs to receive its block rewards; SlushPool and Poolin, for instance, used 56 and 23 distinct wallet addresses, respectively. (b) The counts of inferred MPO transactions; in total, 12,121 transactions were inferred as MPOs' transactions, which corresponds to 0.011\% of the total issued transactions recorded in the Bitcoin blockchain. Poolin has the majority with 2232 (18.41\%), followed by Okex with 2089 (17.24\%) and Huobi with 1666 (13.74\%) transactions. BitDeer and Buffett have the same wallet address as BTC.com and Lubian.com, respectively. We count the addresses of the former as belonging to the latter.}\label{fig:bar-wallet-addresses}
\end{figure}

\subsubsection{Scaling the tests}

While we did not face them in the present work, our test may have two limitations when scaling to large time windows and/or large numbers of transactions.

\begin{table*}[tb]
    \tiny
    \tabcap{Differential prioritization of self-interest transactions}\label{tab:self-interest-txs}
    \resizebox{0.95\textwidth}{!}{
        \begin{tabular}{@{}lccccccc@{}}
            \toprule
            \multicolumn{1}{p{0.8cm}}{\thead{Transactions}} & 
            \multicolumn{1}{l}{\thead{mining pool}} &
            \multicolumn{1}{l}{\thead{norm. hash rate}} &
            \multicolumn{1}{c}{\multirow{2}{*}{$\thead{x}$}} &
            \multicolumn{1}{c}{\multirow{2}{*}{$\thead{y}$}} &
            \multicolumn{2}{c}{\thead{p-value}} &
            \multicolumn{1}{c}{\thead{\% SPPE}} \\
            \multicolumn{1}{c}{\thead{of ...}} &
            \multicolumn{1}{c}{\thead{(m)}} &
            \multicolumn{1}{c}{\thead{($\theta_0$)}} &
            \multicolumn{1}{c}{~} & \multicolumn{1}{c}{~} &
            \multicolumn{1}{c}{\thead{(accel.)}} &
            \multicolumn{1}{c}{\thead{(decel.)}} &
            \multicolumn{1}{c}{\thead{(m)}} \\ 
            \midrule
            {\quad\quad \textit{\textbf{F2Pool}}}
                & F2Pool & 0.1753 & 466 & 839 & \attention{0.0000} & 1.0000 & \attention{78.5494} \\
            \arrayrulecolor{gray}\midrule
            {\quad\quad \textit{\textbf{ViaBTC}}}
                & ViaBTC & 0.0676 & 412 & 720 & \attention{0.0000} & 1.0000 & \attention{98.9175} \\
            \arrayrulecolor{gray}\midrule
            \multirow{2}{*}{\quad\quad \textit{\textbf{1THash \& 58Coin}}}
                & ViaBTC & 0.0676 & 34 & 201 & \attention{0.0000} & 1.0000 & \attention{81.4516} \\
                & 1THash \& 58Coin & 0.0611 & 39 & 201 & \attention{0.0000} & 1.0000 & \attention{96.9143} \\
            \arrayrulecolor{gray}\midrule
            \multirow{2}{*}{\quad\quad \textit{\textbf{SlushPool}}}
                & SlushPool & 0.0375 & 214 & 1343 & \attention{0.0000} & 1.0000 & \attention{88.3082} \\
                & ViaBTC & 0.0676 & 140 & 1343 & \attention{0.0000} & 1.0000 & \attention{45.1523} \\
            \arrayrulecolor{black}\bottomrule
        \end{tabular}
    } 
\end{table*}

First, it may become difficult to compute the $p$-value from the binomial distribution for large values of $y$. In such cases, we can use the following approximation for our analysis: If $y$ is large enough and $\theta_0$ is not close to zero or one (i.e., $x$ and $y-x$ are large enough), the binomial distribution of parameters $\theta_0$ and $y$ is well approximated by the normal distribution with mean $y\theta_0$ and variance $y\theta_0(1-\theta_0)$. Hence, the $p$-value for the acceleration test can be computed as,
\begin{align*}
  p \simeq \Phi \left( \frac{x-y\theta_0}{\sqrt{y\theta_0(1-\theta_0)}} \right),
\end{align*}
where $\Phi$ is the CDF of a standard normal random variable. A similar approximation can be done for the deceleration test.

Second, the hash rates of miners in our $p$-value test are assumed to be more or less constant (i.e., $\theta_0$ is a constant). This assumption is a limitation of our test as, in reality, hash rates of miners may vary over time, particularly over large time windows. In such situations, our test results may be affected, particularly when the arrival times of transactions are not regularly spread over the time window of our analysis. We address this issue in the current paper by confirming the results of the $p$-value test through the SPPE-test, which is not affected by variable hash rates. It is possible, however, to alleviate this limitation of our analysis. One natural way is to divide the total time window into multiple windows such that the hash rate is more or less constant in those shorter time windows; and compute $p$-values in each time window. We can then combine the obtained $p$-values using Fisher's method \cite{Fisher@1992Statistical,Fisher@1948}. We leave the investigation of such extended test procedures to future work, when they might be needed.

\subsection{Self-interest transactions}

\begin{table*}[tb]
    \begin{center}
        \tiny
        \tabcap{Differential prioritization of scam-payment transactions}\label{tab:twitter-scam-txs}
        \resizebox{0.75\textwidth}{!}{%
            \begin{tabular}{@{}ccccccr@{}}
                \toprule
                \multicolumn{1}{c}{\thead{mining pool}} &
                \multicolumn{1}{l}{\thead{norm. hash rate}} &
                \multicolumn{1}{c}{\multirow{2}{*}{$\thead{x}$}} &
                \multicolumn{1}{c}{\multirow{2}{*}{$\thead{y}$}} &
                \multicolumn{2}{c}{\thead{p-value}} &
                \multicolumn{1}{c}{\thead{\% SPPE}} \\
                \multicolumn{1}{c}{\thead{(m)}} &
                \multicolumn{1}{c}{\thead{($\theta_0$)}} &
                \multicolumn{1}{c}{~} & \multicolumn{1}{c}{~} &
                \multicolumn{1}{c}{\thead{(accel.)}} &
                \multicolumn{1}{c}{\thead{(decel.)}} &
                \multicolumn{1}{c}{\thead{(m)}} \\ 
                \midrule
                Poolin & 0.1528 & 10 & 53 & 0.2856 & 0.8227 & 	$-3.9787$ \\
                F2Pool & 0.1450 & 10 & 53 & 0.2323  & 0.8629 & $0.8735$ \\
                BTC.com & 0.1147 & 9  & 53 & 0.1483  & 0.9233 & $-2.8333$ \\
                AntPool & 0.1093 & 4  & 53 & 0.8450 & 0.2989  & $31.5000$  \\
                Huobi  & 0.0955 & 1  & 53 & 0.9951 & 0.0323 &  $-1.6428$ \\
                Okex & 0.0698 & 3  & 53 & 0.7248 & 0.4890  & $-5.0000$  \\
                1THash \& 58COIN & 0.0684 & 8  & 53 & 0.0268  & 0.9907  & $-0.5000$ \\
                Binance Pool   & 0.0590 & 3  & 53 & 0.6120 & 0.6180  & $-2.6000$ \\
                ViaBTC & 0.0552 & 1  & 53 & 0.9507 & 0.2020 &  $-4.0000$ \\ \bottomrule
            \end{tabular}%
        } 
    \end{center}
\end{table*}

To identify transactions where a mining pool is a sender or receiver of transactions, we first need to identify Bitcoin wallets (addresses) that belong to mining pools.
In Bitcoin, whenever a mining pool discovers a new block, it specifies a wallet address to receive the mining rewards. 
This mining pool address is included in the Coinbase transaction (refer~\S\ref{sec:background}) that appears at the start of every block.
In our data set \dsc{}, we gathered all the wallet addresses used by the top-$20$ mining pools to receive their rewards.
For each mining pool, we then retrieved all committed transactions, in which coins were sent from the mining pool's wallet.
Figure~\ref{fig:bar-wallet-addresses} shows the statistics for the mining pool wallets and the transactions spending (sending) coins from (to) the wallets, for each of the top-$20$ mining pools in data set \dsc{}.
%
%
%
We found hundreds or thousands of self-interest transactions for most of the mining pools. 

\subsubsection{Acceleration of self-interest transactions}

For self-interest transactions belonging to each of the top-20 mining pools, we separately applied our statistical test to check whether any of the top-10 mining pools (that mined at least $4\%$ of all mined blocks in data set \dsc{}) are preferentially accelerating or decelerating the transactions.
In Table~\ref{tab:self-interest-txs}, we report the statistics from our test for mining pools that were found to preferentially treat transactions belonging to their own or other mining pools.
Strikingly, Table~\ref{tab:self-interest-txs} shows that 4 out of the top-10 mining pools namely, F2Pool, ViaBTC, 1THash \& 58Coin, and SlushPool \stress{selfishly accelerated} their own transactions, i.e., coin transfers from or to their own accounts (p-value for acceleration test is less than $0.001$).
Equally, if not more interestingly, Table~\ref{tab:self-interest-txs} shows collusive behavior among mining pools. 
Specifically, it shows that transactions issued by 1THash \& 58Coin and SlushPool were \stress{collusively accelerated} by ViaBTC (p-value for acceleration test is less than $0.001$).
That these mining pools were accelerating the transactions is further confirmed by the SPPE measure, which clearly shows that in each of the above cases, the self-interest transactions were also being included within the blocks ahead of other higher fee-rate transactions.

\subsection{Scam-payment transactions}


Next, we investigate whether any mining pool attempted to decelerate or exclude scam-payment transactions.  

On July 15, 2020, multiple celebrities'
accounts on Twitter fell prey to a scam attack.
The scammers posted the message that anyone who transferred bitcoins to a specific
wallet will receive twice the amount in return~\cite{Frenkel@nyt20}.
In response, several people sent, in total, $12.87051731$ bitcoins---then
worth nearly $\num{142000}$ (USD)---to the attacker's wallet via $386$
transactions, which were confirmed across $53$ blocks by $12$ miners.

To examine the miners' behavior during this scam attack, we selected all blocks mined from July 14 to August 9, 2020 (i.e., \num{3697} blocks in total, containing \num{8318621} issued transactions as described in~\S\ref{sec:supp-scam-txs}) from our data set \dsc.
%
%
Once again, we applied our statistical test to check whether any of the top-$9$ mining pools (that mined at least $5\%$ of all mined blocks from this data) are preferentially accelerating or decelerating the transactions.
Table~\ref{tab:twitter-scam-txs} shows the test statistics.
Interestingly, we find no statistically significant evidence (i.e., p-value less than $0.001$) of scam-payment acceleration or deceleration across all top mining pools. 
Looking at SPPE measure across the mining pools, we find no evidence of mining pools (other than AntPool) preferentially ordering the scam-payment transactions within blocks.
In short, our findings show that most mining pool operators today do not distinguish between normal and scam-payment transactions.

\subsection{Dark-fee transactions}
\label{sec:dark-fee-txs}

We refer to transactions that offer additional fees to specific mining pools through an opaque and non-public side-channel payment as dark-fee transactions.
Many large mining pool operators allow such side-channel payments on their websites for users wanting to ``accelerate'' the confirmation of their transactions, especially during periods of congestion.
%
%
Such private side-channel payments that hide the fees a user pays to miners from others have other benefits for the users~\cite{BTC@accelerator,Taichi@accelerator,F2Pool@accelerator,Poolin@accelerator,AntPool@accelerator}. One well-known advantage is, for instance, avoiding the fee-rate competition in transaction inclusion, particularly during periods of high \mpool congestion; private side-channel payments would reduce a user's transaction cost volatility and curb front-running risks~\cite{Daian@S&P20,strehle2020exclusive,Eskandari@FC-2020}.
We use the data set \dsc{} to first investigate how such transaction acceleration services work and later propose a simple test for detecting accelerated transactions in the Bitcoin blockchain.

\subsubsection{Investigating transaction acceleration services}

We examined transaction acceleration services offered by $5$ large Bitcoin mining pools namely, BTC.com~\cite{BTC@accelerator}, AntPool~\cite{AntPool@accelerator}, ViaBTC~\cite{ViaBTC@accelerator}, F2Pool~\cite{F2Pool@accelerator}, and Poolin~\cite{Poolin@accelerator}.
Specifically, we queried BTC.com for the prices of accelerating all transactions in a real-time snapshot of the \mpool in data set \dsc (see~\S\ref{sec:tx-accelerator-comparison}). %
We found that the dark fee requested by BTC.com to accelerate each transaction is so high that if it was added to the publicly offered transaction fee, the resulting total fee-rate would be higher than the fee-rate offered by any other transaction in the \mpool snapshot.
Put differently, had users included the requested acceleration fees in the publicly offered fee when issuing the transaction, every miner would have included the transaction with the highest priority.

The above observation raises the following question: \stress{why would rational users offer a dark fee to incentivize a subset of miners to prioritize their transaction rather than publicly announce the fee to incentivize all miners to prioritize their transaction?}
One potential explanation could be that as payment senders determine the publicly offered transaction fees, payment receivers might wish to accelerate the transaction confirmation by offering an acceleration fee. 
Another explanation could be that the user issuing the transaction might want to avoid revealing the true fees they are willing to offer publicly, to avoid a fee-rate battle with transactions competing for inclusion in the chain during congestion. Opaque transaction fees can reduce transaction cost volatility, but they may also unfairly bias the level playing field amongst user transactions attempting to front-run one another~\cite{strehle2020exclusive,Daian@S&P20}.   

On the other hand, every rational mining pool has clear incentives to offer such acceleration services.
They receive a very high fee by mining the accelerated transaction. 
Better still, they keep the offered fee, even if the accelerated transaction were mined by some other miners. 

\subsubsection{Detecting accelerated transactions}

\begin{table}[tb]
    \small
    \begin{center}
        \tabcap{For an SPPE $\ge$ 99\%, we observe that 64.98\% of BTC.com transactions were accelerated; the fourth column values are derived by dividing the values in the second with those in the third. The number of accelerated transactions decreases to 18.12\% for an SPPE $\ge$ 90\% and to 1.06\% for an SPPE $\ge$ 50\%.}\label{tab:sppe-tx-violation-acceleration}
        \resizebox{.45\textwidth}{!}{%
            \begin{tabular}{rrrr}
            \toprule
            \multicolumn{1}{c}{\thead{SPPE ($\ge$)}} & \thead{\# txs} & \thead{\# acc. txs} & \thead{\% acc. txs} \\ \midrule
            $100\%$                     & \num{628}     & \num{464}     & $73.89$       \\
            $99\%$                      & \num{1108}     & \num{720}     & $64.98$       \\
            $90\%$                      & \num{5365}   & \num{972}      & $18.12$       \\
            $50\%$                      & \num{95282}  & \num{1007}      & $1.06$          \\ 
            $1\%$                       & \num{657423}  & \num{1029}      & $0.16$          \\ 
            \bottomrule
            \end{tabular}
        } 
    \end{center}
\end{table}

Given the high fees demanded by acceleration services, we anticipate that \stress{accelerated transactions would be included in the blockchain with the highest priority}, i.e., in the first few blocks mined by the accelerating miner and amongst the first few positions within the block. 
We would also anticipate that \stress{without the acceleration fee, the transaction would not stand a chance of being included in the block based on its publicly offered transaction fee}.
The above two observations suggest a potential method for detecting accelerated transactions in the Bitcoin blockchain:
An accelerated transaction would have a very high \textit{\textbf{signed position prediction error (SPPE)}}, as its predicted position based on its public fee would be towards the bottom of the block it is included in, while its actual position would be towards the very top of the block. 

To test the effectiveness of our method, we analyzed all \num{6381} blocks and \num{13395079} transactions mined by BTC.com mining pool in data set \dsc{}. 
%
We then extracted all transactions with SPPE greater or equal than $100\%$, $99\%$, $90\%$, $50\%$, $1\%$ and checked what fraction of such transactions were accelerated.
%
Given a transaction identifier, BTC.com's acceleration service~\cite{BTC@accelerator} allows anyone to verify whether the transaction has been accelerated. 
Our results are shown in Table~\ref{tab:sppe-tx-violation-acceleration}. 
We find that more than $64\%$ of the \num{1108} transactions with SPPE greater or equal than $99\%$ were accelerated, while only $1.06\%$ of transactions with SPPE greater or equal than $50\%$ were accelerated.
In comparison, we found no accelerated transactions in a random sample of \num{1000} transactions drawn from the \num{13395079} transactions mined by BTC.com.
Our results show that large values of SPPE for confirmed transactions indicate the potential use of transaction acceleration services.
In particular, a transaction with SPPE $\ge 99\%$ (i.e., a transaction that is included in the top $1\%$ of the block positions, when it should have been included in the bottom $1\%$ of the block positions based on their public fee-rate) has a high chance of being accelerated.

\section{Concluding Discussion}\label{sec:conclusion}

At a high-level, our analysis of transaction ordering in the Bitcoin blockchain offers three important takeaways.

\begin{enumerate}
\item {\it Selfish transaction prioritization:} We showed that miners do not fully follow the expected fee-rate based prioritization norms in Bitcoin, especially for transactions where they have a vested (selfish) interest. 
\item {\it Dark-fee transaction prioritization:} We demonstrated that not all fees offered by transactions are transparent and public. Miners can accept so-called ``dark-fee'' payments via side channels to accelerate transactions. 
\item {\it Collusive transaction prioritization:} We showed that miners collude on accelerating transactions in which other miners have a vested interest. 
\end{enumerate}

While the percentage of transactions that are affected by selfish, non-transparent, and collusive behaviors of miners is relatively small today, if unchecked, the spreading of such misbehaviors portends serious trouble for future blockchain systems.
The transaction fees offered by Bitcoin users during periods of congestion crucially relies, for instance, on the assumption that the total fees offered by other transactions are public and transparent. 
If some transactions offer opaque (or dark) fees, it becomes hard for Bitcoin users wishing to get their transactions confirmed before a deadline to offer the correct fee and have their transaction accepted.
Similarly, miners receiving dark fees have a clear, unfair advantage over other miners, as they receive higher fees for mining the same transaction. 
Worse, the dark fee receiving miners get to keep the additional fee, even when the transaction is mined by other miners. 
Finally, collusion between mining pools further concentrates the activities of the whole network to the hands of a few large mining pools.   

Since the mechanism for prioritizing transactions is similar across most popular cryptocurrencies~\cite{strehle2020exclusive,Daian@S&P20,Roughgarden@EC21}, our methodology to study miners' adherence can be generalized to other blockchains (e.g., Ethereum).

\subsection{The Case for Chain Neutrality}

Our findings call for a community-wide debate on defining transaction prioritization norms and enforcing them in a transparent manner. Specifically, we highlight three challenging questions that need to be addressed for the future.
%

$\star{}$
{\it What are the desired transaction prioritization norms in public \pow{}
blockchains?}
What aspects of transactions besides fee-rate should miners be allowed to consider when ordering them? For instance, should the waiting time of transactions also be considered to avoid indefinitely delaying some transactions? Should the transaction value (i.e., amount of bitcoins transferred between different accounts) be a factor in ordering, as fee-rate based ordering favors larger value over smaller value transactions? Similarly, while we did not find evidence of miners decelerating or censoring (i.e., refusing to mine) transactions, the current protocols do not disallow such discriminatory behaviors by miners. Should prioritization norms also explicitly disallow discriminating transactions based on certain transaction features like sending or receiving wallet addresses? Such norms would be analogous to {\it network neutrality} norms for ISPs that disallow flows from being treated differently based on their source/destination addresses or payload.  

$\star{}$
{\it How can we ensure that the distributed miners are adhering to desired and defined norms?} 
Miners in public PoW blockchains operate in a distributed manner, over
a P2P network.
This model of operation results in different miners potentially having distinct, 
typically different, views of the state of the system (e.g., set of outstanding transactions).
Given these differences, are there mechanisms (say, based on
statistical tests~\cite{lev2020fairledger,Orda2019,Asayag18a}) that
any third-party observer could use to verify that a miner adheres to the
established norm(s)?

$\star{}$
{\it How can we model and analyze the impact of selfish, non-transparent, collusive behaviors of miners?}
While the above themes align well with a long-term vision of defining
and enforcing well-defined ordering norms in blockchains, in the short term one could focus on examining the implications of the norm violations in today's blockchains.
Specifically, how can we characterize the ordering that would
result from different miners following different prioritization norms,
especially given an estimate of miners' hashing or mining powers (i.e., their
likelihood of mining a block).
Such characterization has crucial implications for Bitcoin users.





\section*{Acknowledgments}\label{sec:ack}

K. P. Gummadi acknowledges support from the European Research Council (ERC) Advanced Grant ``Foundations for Fair Social Computing,'' funded under the European Union's Horizon 2020 Framework Programme (grant agreement no. 789373).
P. Loiseau was supported by MIAI @ Grenoble Alpes (ANR-19-P3IA-0003) and by the French National Research Agency through grant ANR-20-CE23-0007.
A. Mislove acknowledges support from NSF grants CNS-1900879 and CNS-1955227.

J. Messias dedicates this work to his late father, Jota Missias~\cite{JotaMissias@Wiki}.

%



\bibliographystyle{ACM-Reference-Format}
\bibliography{references}

\appendix
\section{Congestion in \mpool{} of \dsb{}} \label{sec:supp-tx-ord}

\begin{figure}[h]
    \centering
    \includegraphics[width={\onecolgrid}]{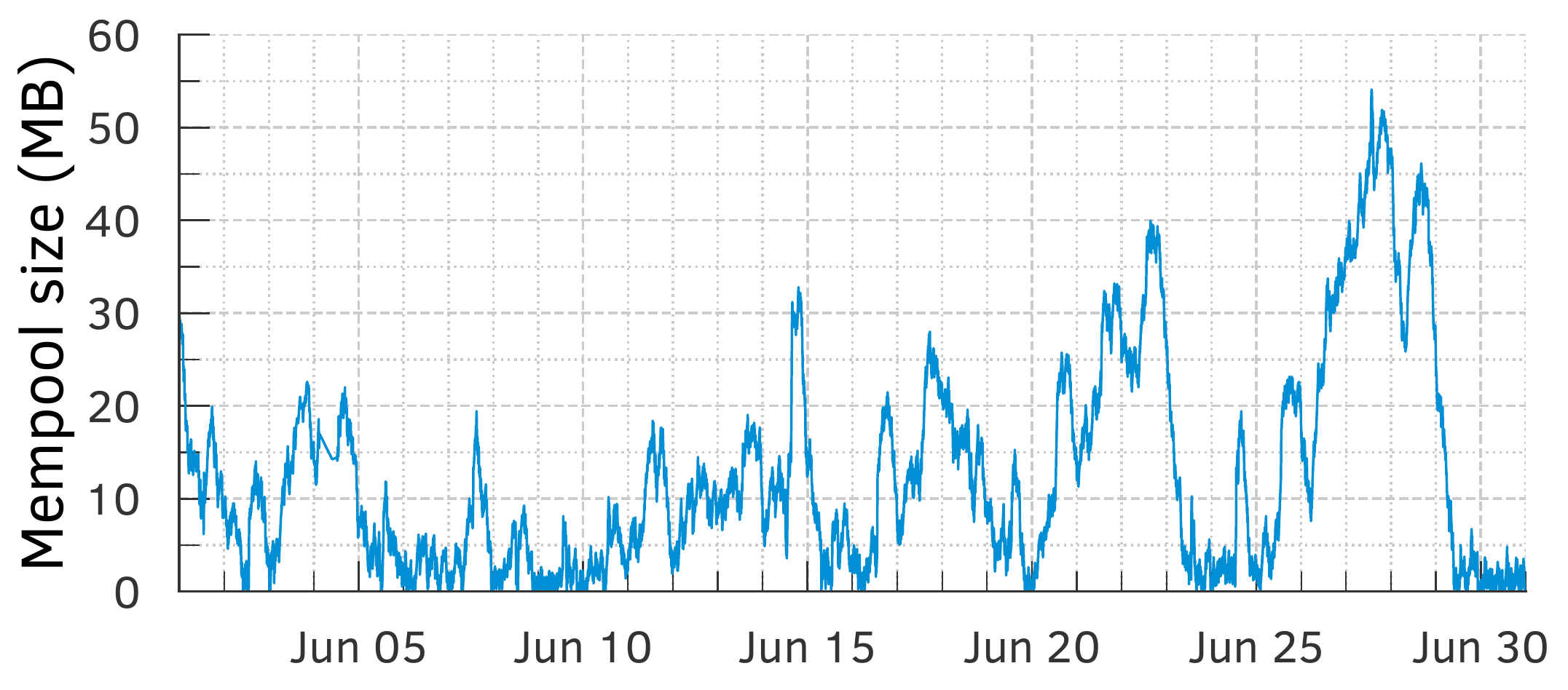}
    \figcap{\mpool{} size from \dsb{} as a function of time.}\label{fig:mpool-sz-b}
\end{figure}

Congestion in \mpool{} is typical not only in \dsa{} (as discussed in~\S\ref{subsec:cong-delays}), but also in \dsb{}.
Indeed, Figure~\ref{fig:mpool-sz-b} reveals a huge variance in \mpool{} congestion,
much higher than that observed in \dsa{}.
\mpool{} size fluctuations in \dsb{} are, for instance, approximately three times
higher than that in \dsa{}.
Around June $22\tsup{nd}$, there was a surge in Bitcoin price
following the announcements of Facebook’s Libra\footnote{On June 18\tsup{th}, Facebook
announced its cryptocurrency, Libra, which was later renamed to Diem.~\url{https://www.diem.com}} and another surge
around June $25\tsup{th}$ after the news of US dollar
depreciation~\cite{CNN-BITCOIN-2019}.
These price surges significantly increased the number of transaction issued, which in turn introduced delays.
As a consequence, at times, \mpool{} in \dsb{} takes much longer duration than
in \dsa{} to be drained of all transactions.

\section{Significance of Transaction Fees} \label{sec:signif-tx-fees}

Table~\ref{tab:fee-revenue} shows the contribution of transaction fees towards miners' revenue across all blocks mined from 2016 to 2020. In 2018, fees accounted for an average of $3.19\%$ of miners' total revenue per block; in 2019 and 2020 were $2.75\%$ and $6.29\%$, respectively. However, if we consider only blocks mined from May 2020 (i.e., blocks with a mining reward of $6.25$ BTC), the fees account for, on average, $8.90\%$ with an std. of $6.54\%$ in total. Therefore, revenue from transaction fees is increasing~\cite{Easley-SSRN2017}, and it tends to continue.

\begin{table}[h]
    \begin{center}
    \tabcap{Miners' relative revenue from transaction fees (expressed as a percentage of the total revenue) across all blocks mined from 2016 until the end of 2020.}\label{tab:fee-revenue}
    \resizebox{.45\textwidth}{!}{%
        \begin{tabular}{rcrccccrc}
        \toprule
        \multicolumn{1}{c}{\multirow{1}{*}{\thead{Year}}} & \multicolumn{1}{c}{\thead{\# of blocks}} & \multicolumn{1}{c}{\thead{mean}} & \multicolumn{1}{c}{\thead{std}} & \multicolumn{1}{c}{\thead{min}} & \multicolumn{1}{c}{\thead{25-perc}} & \multicolumn{1}{c}{\thead{median}} & \multicolumn{1}{c}{\thead{75-perc}} & \multicolumn{1}{c}{\thead{max}}\\  \midrule
            2016 & \num{54851}        & 2.48  & 2.12 & 0    & 0.87    & 1.78   & 3.84    & 92.10 \\
            2017 & \num{55928}        & 11.77 & 7.73 & 0    & 6.33    & 10.49  & 15.58   & 86.44 \\
            2018 & \num{54498}        & 3.19  & 5.85 & 0    & 0.52    & 1.22   & 2.60    & 44.19 \\
            2019 & \num{54232}        & 2.75  & 2.77 & 0    & 0.80    & 1.81   & 3.70    & 24.32 \\
            2020 & \num{53211}        & 6.29  & 6.34 & 0    & 1.37    & 4.00   & 9.71    & 39.46 \\
        \bottomrule
        \end{tabular}
    } 
    \end{center}
\end{table}

\begin{figure}[b]
    \centering
    \includegraphics[width={\onecolgrid}]{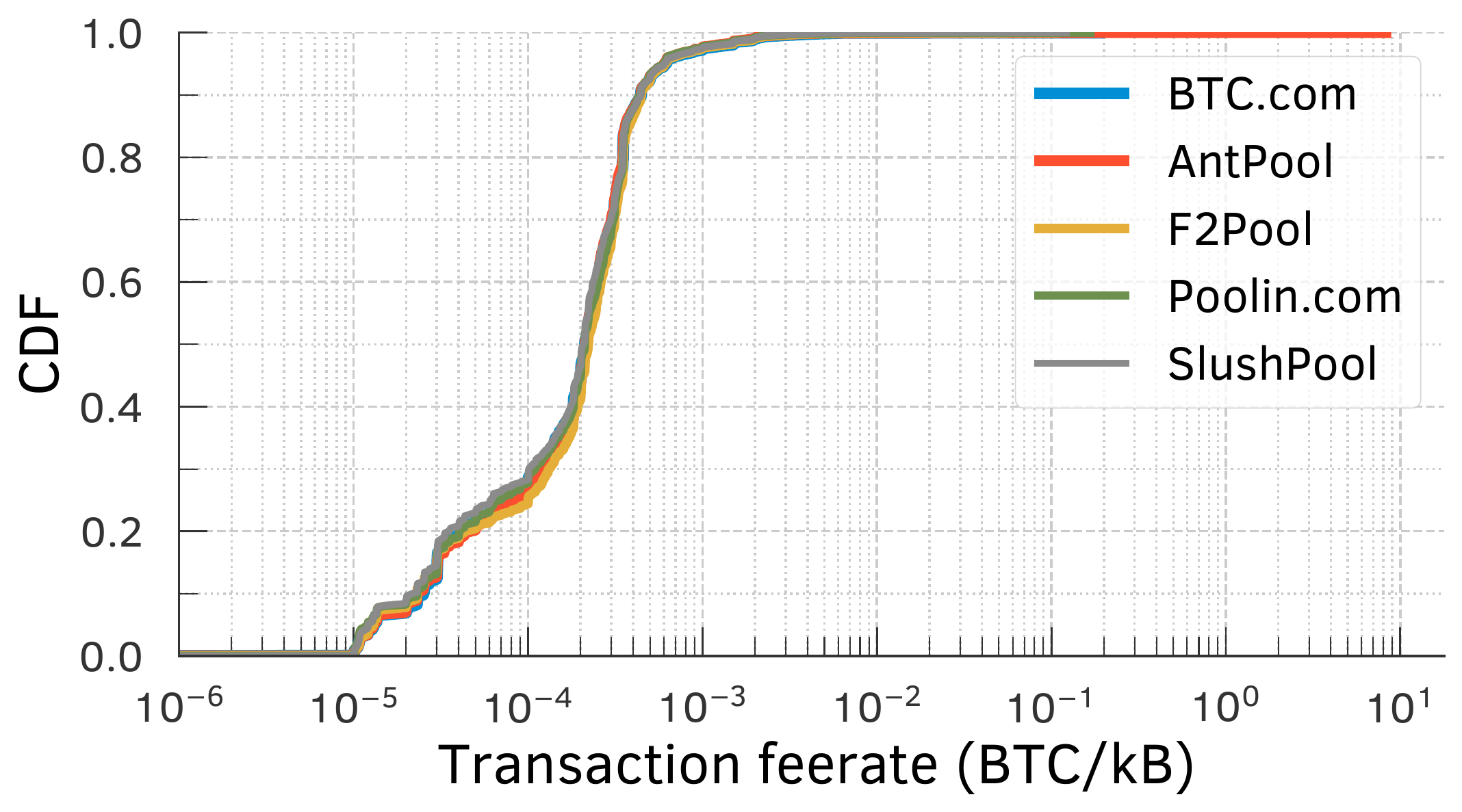}
    \figcap{Distributions of fee-rates for transactions committed by the top-5 mining pools in data set \dsa.}\label{fig:cdf-fee-top5}
\end{figure}

\section{Transaction Fee-rates across MPOs} \label{sec:tx-fees-across-mpos}

Transaction fee-rate of committed transactions in both data sets \dsa{} and \dsb{} exhibits a wide range, from
$10^{-6}$ to beyond $\uTxFee{1}$.
A comparison of the fee-rates of transactions in \dsa{} committed by the top
five mining pool operators (in a rank ordering of mining pool operators based on
the number of blocks mined), in Figure~\ref{fig:cdf-fee-top5}, shows no major
differences in fee-rate distributions across the different MPOs. Around $70\%$ of the transactions offer from $10^{-4}$ to $10^{-3}$ \uTxFee{} that is one to two orders of magnitude more than the recommended minimum of $10^{-5}$ \uTxFee{}. We hypothesize that users increase the fee-rates offered during high \mpool congestion---they assume that higher the fee-rate implies lower the transaction delay or commit time.

\section{On Fee-rates and Congestion} \label{sec:fees-and-cong}


\begin{figure}[t]
    \centering
    \includegraphics[width={\onecolgrid}]{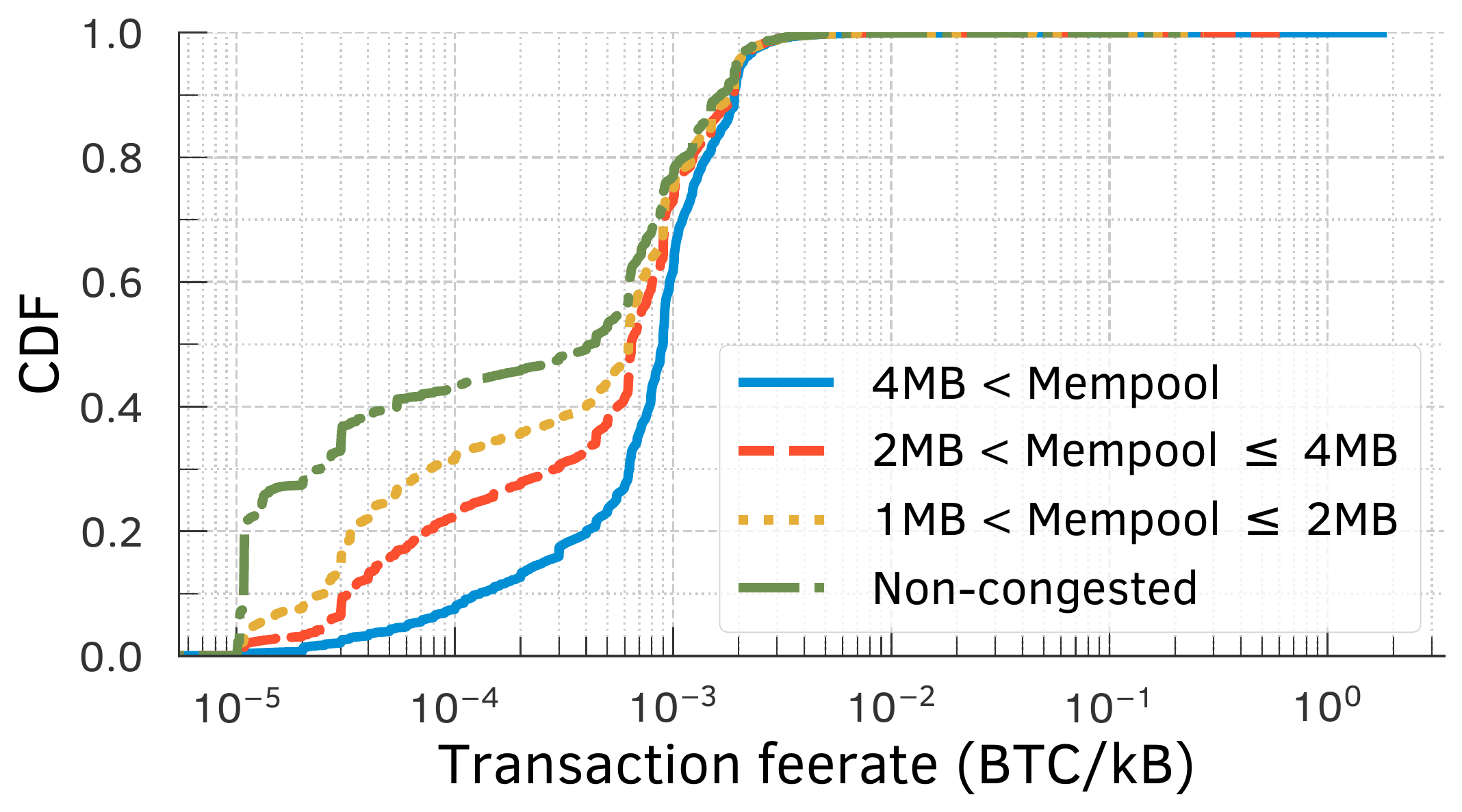}
    \figcap{Distributions of transaction-commit delays for transactions in \dsb{} issued at different congestion levels.}\label{fig:fee-cong-rel-b}
\end{figure}

In Figure~\ref{fig:fee-cong-rel-b}, we show the fee-rates of transactions observed in $4$ different bins or congestion levels in data set \dsb.
Each bin in the plot corresponds to a specific level of congestion identified by the \mpool size:
lower than \uMB{1} (\stress{no congestion}), in $(1, 2]$ MB (\stress{lowest congestion}), in $(2, 4]$ MB, and higher than \uMB{4} (\stress{highest congestion}).
Fee rates at high congestion levels are strictly higher (in distribution, and hence also on average) than those at low congestion levels.
Users, therefore, increase transaction fees to mitigate the delays incurred during congestion.

\begin{figure}[h]
    \centering
    \includegraphics[width={\onecolgrid}]{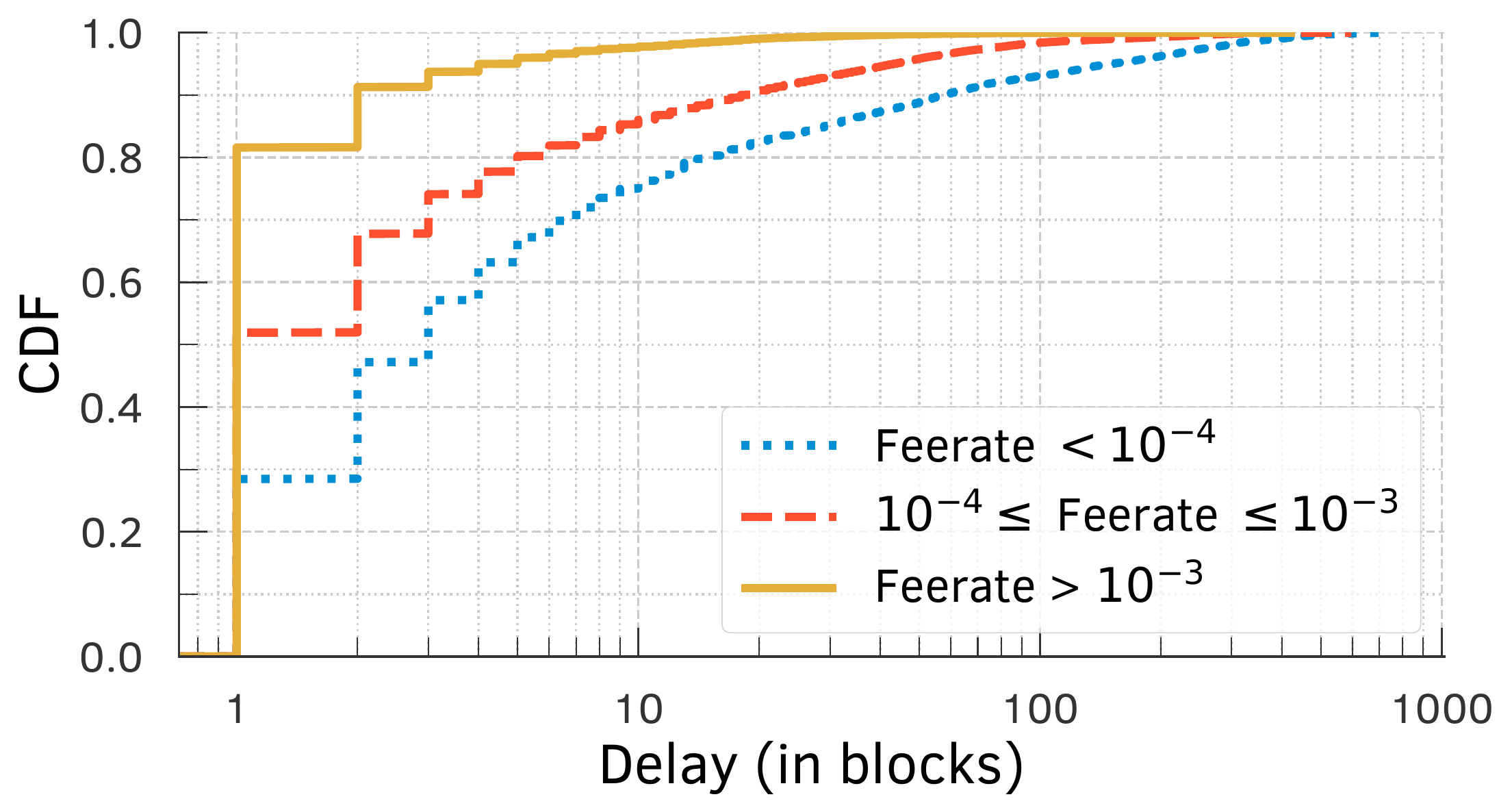}
    \figcap{Distributions of transaction-commit delays in \dsb{} for different transaction fee-rates.}\label{fig:fee-delay-rel-b}
\end{figure}

Figure~\ref{fig:fee-delay-rel-b} shows that users' strategy of increasing
fee-rates to combat congestion seems to work well in practice---higher the fee rate, lower the transaction commit delay.
Here, we compare the CDF of commit delays of transactions with low (i.e., less
than $10^{-4}$~\feeunit{}), high (i.e., between $10^{-4}$ and
$10^{-3}$~\feeunit{}), and exorbitant (i.e., more than $10^{-3}$) fee-rates, in data set \dsb.
The commit delays for transactions with high fee-rates (i.e., greater than $10^{-3}$~\feeunit{}) are significantly smaller than those with low fee-rates (i.e., lesser than $10^{-4}$~\feeunit{}).

\section{Child-Pays-For-Parent Transactions} \label{sec:cpfp-txs}

Given any block $B_i$ that contains a set of issued transactions $T = \{t_0, t_1, \cdots, t_n\}$, where each transaction has at least one transaction input identifier $V = \{v_0, v_1, \cdots, v_m\}$, the transaction $t_j \in T$ is said to be a \newterm{child-pays-for-parent transaction (CPFP-tx)} if and only if there exists at least one input $v_k \in V$ that belongs to $T$. In other words, a transaction is a CPFP-tx if and only if it spends from any previous transaction that was also included in the same block $B_i$.

\begin{figure}[h]
  \centering
  \includegraphics[width={\onecolgrid}]{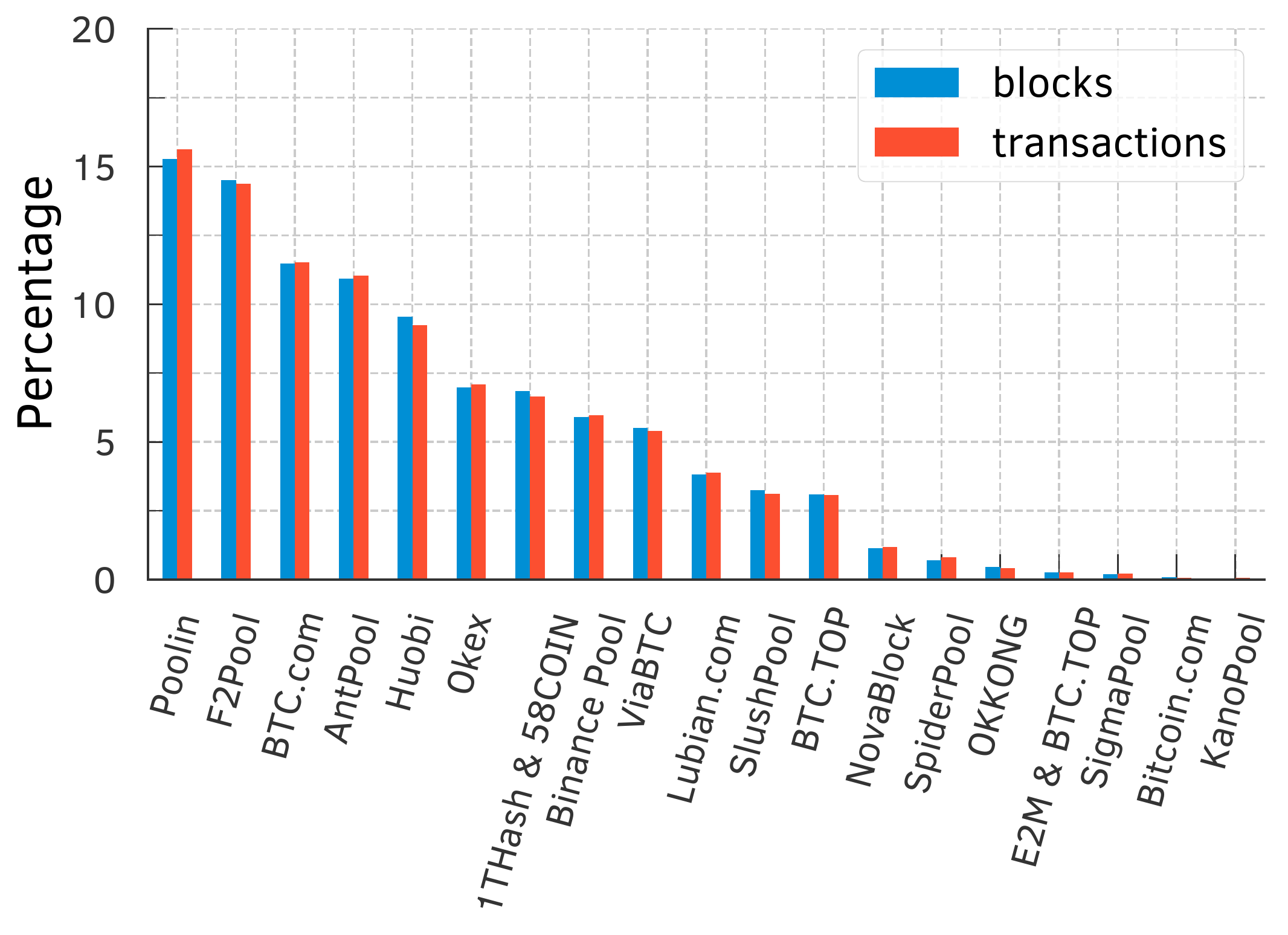}
  \figcap{Distribution of blocks mined and transactions confirmed by different MPOs during the Twitter Scam attack from July 14\tsup{th} to August 9\tsup{th}, 2020.}\label{fig:dist-txs-blks-twitter}
\end{figure}

\section{Miners' Behavior During the Scam} \label{sec:supp-scam-txs}

To examine the miners' behavior during the Twitter scam attack from July 14\tsup{th} to August 9\tsup{th}, 2020, we selected  all blocks
mined (\num{3697} in total, containing \num{8318621} issued transactions) during this time period from our data set \dsc.
If we rank the MPOs responsible for these blocks by the number of blocks ($B$) mined (or, essentially, the
approximate hashing capacity $h$), the top five MPOs (refer Figure~\ref{fig:dist-txs-blks-twitter}) turn out to be Poolin ($B$: $\num{565}$; $h$: $15.28\%$), F2Pool ($B$: $\num{536}$; $h$: $14.5\%$), 
BTC.com ($B$: $\num{424}$; $h$: $11.47\%$), AntPool ($B$: $\num{404}$; $h$: $10.93\%$), and Huobi ($B$: $\num{353}$; $h$: $9.55\%$).

\begin{figure}[tb]
  \centering
  \includegraphics[width={\onecolgrid}]{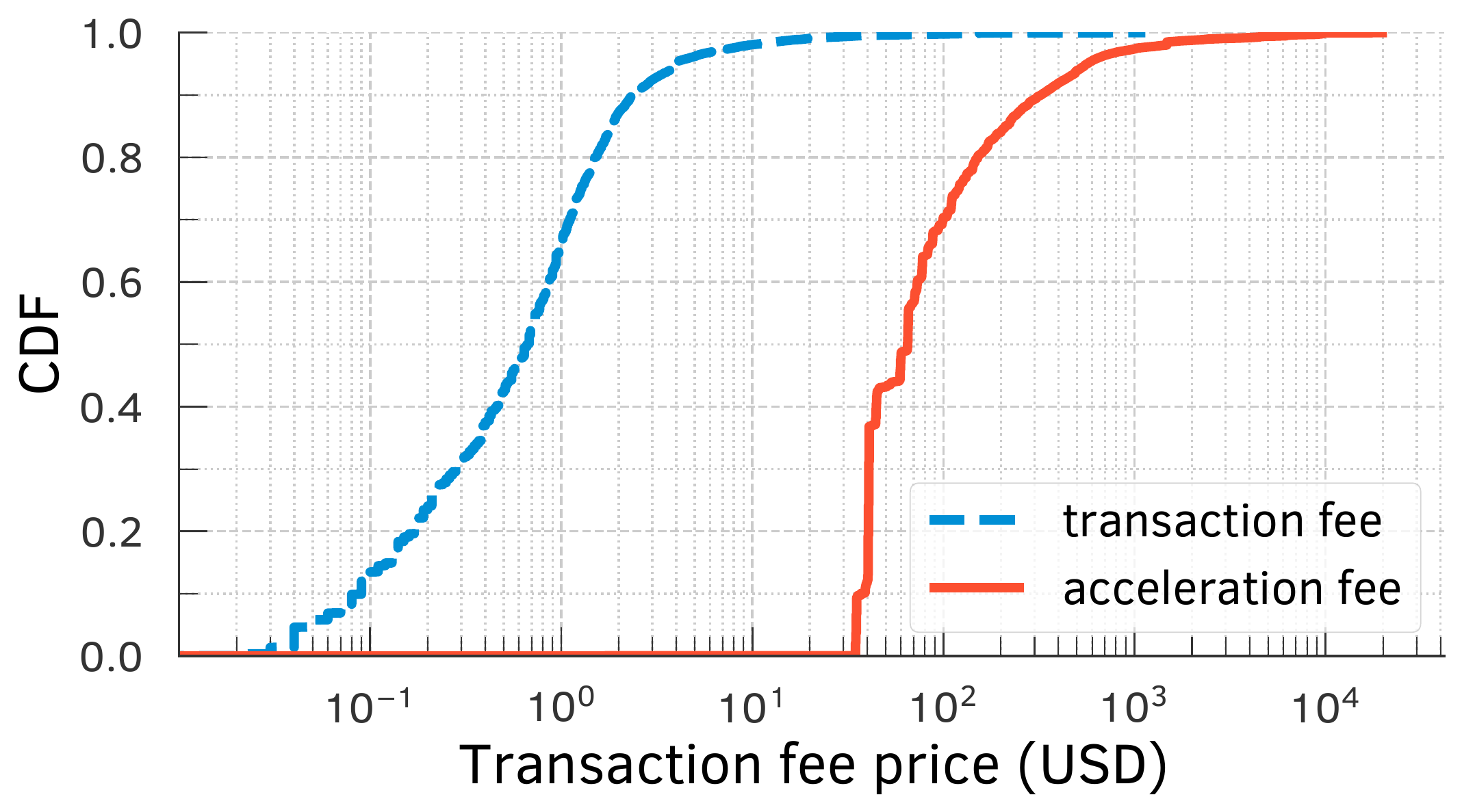}
  \figcap{Fee price comparison between the transaction fee and the acceleration services from an snapshot of our \mpool on November 24\tsup{th}, 2020. Acceleration service provided by BTC.com is on average 566.3 times higher (4734.67 of std.) and on median 116.64 times higher than the Bitcoin transaction fees. The minimum is 0.54, the 25-perc is 51.64, and the 75-perc and the maximum are 351.8 and 428,800, respectively.}\label{fig:acceleration-fee-price-comparison}
\end{figure}

\section{Transaction-Acceleration Fees}
\label{sec:tx-accelerator-comparison}

In this experiment, we compare the transaction-acceleration fee with the typical transaction fees in Bitcoin.
To this end, we retrieved a snapshot containing \num{26332} unconfirmed transactions from our node's \mpool on November 24\tsup{th} 2020 at 10:08:41 UTC.
Then, for each transaction, we searched its respective transaction accelerator price (or acceleration fee) via the acceleration service provided by BTC.com~\cite{BTC@accelerator}.
We inferred the acceleration fees for \num{23341} ($88.64\%$) out of the \num{26332} unconfirmed transactions.
Figure~\ref{fig:acceleration-fee-price-comparison} shows the CDF of both the Bitcoin transaction fees as well as the acceleration fees provided by BTC.com.
Acceleration fee is on average $566.3$ times higher (\num{4734.67} of std.) and on median 116.64 times higher than the Bitcoin transaction fees.
At the time of this experiment, 1~BTC was worth \num{18875.10}~USD.

\end{document}